\documentclass[%
reprint,
aps,
prb,
]{revtex4-2}

\usepackage{graphicx}
\usepackage{hyperref}
\usepackage{dcolumn}
\usepackage{bm}
\usepackage{amsmath}
\usepackage{natbib}
\usepackage{booktabs}
\DeclareMathAlphabet{\mathbit}{OT1}{cmr}{bx}{it}

\begin{document}

\title{Theory of proximity effect in two-dimensional unconventional 
superconductor with Rashba spin-orbit interaction}

\author{Shun Tamura$^{1}$, Yukio Tanaka$^{1}$}
\affiliation{%
   $^1$~Department of Applied Physics, Nagoya University, Nagoya 464--8603, Japan\\
}%

\begin{abstract}
   We study the anomalous proximity effect in diffusive normal metal 
   (DN)/unconventional superconductor junctions, where the local density of states (LDOS) in the DN has a 
   zero-energy peak due to the penetration of the odd-frequency spin-triplet 
   $s$-wave pairing. 
   In this study, we consider a two-dimensional unconventional 
   superconductor on the substrate in the presence of a 
   Rashba spin-orbit coupling (RSOC) $\lambda$, where the Rashba vector 
   is parallel to the $z$-direction.
   The anomalous proximity effect, originally predicted in 
   spin-triplet $p$-wave superconductor junctions, is sensitive 
   to the RSOC\@.
   It disappears with the increase of $\lambda$.
   On the other hand, the anomalous proximity effect can be switched on by 
   the large $\lambda$
   values in the spin-singlet $d_{xy}$-wave superconductor junctions. 
   The resulting zero-energy LDOS and the magnitude of the odd-frequency 
   spin-triplet $s$-wave pair amplitude increase with the increase of $\lambda$.

\end{abstract}
\pacs{pacs}

\maketitle
\thispagestyle{empty}


\section{Introduction}

In unconventional superconductors (SCs), 
nodes of the energy gap and the change of the sign of the 
pair potential (gap function) are generated on the Fermi surface. 
It is known that the change of the sign of the pair potential produces 
in-gap zero-energy states localized near the surface, known as 
zero-energy surface Andreev bound states (ZESABSs)\cite{ABS,ABSb,Hu94,ABSR1,ABSR2}. 
The ZESABS manifests itself as a 
zero bias conductance peak in quasiparticle tunneling 
experiments~\cite{TK95,ABSR1,Experiment3,Experiment4,Experiment5}.

Nowadays, it is recognized that 
the ZESABS has deep physical meanings from the aspects of 
topology and symmetry in condensed matter physics~\cite{tanaka12}. 
It is known that the ZESABS is protected by the topological invariant defined in the
bulk Hamiltonian\cite{RH02,schnyder08,STYY11,tanaka12,SatoFujimoto2016,SatoAndo2017}. 
Thus, unconventional superconductors hosting nodes can be classified as 
topological superconductors\cite{Sato06,STYY11,Kobayashi2014}. 
Besides this property, the ZESABS has a significant meaning from the aspect of the
symmetry of Cooper pairs. 
\par

The breaking of translational invariance induces an odd-frequency pairing, 
$i.e$., spin-singlet odd-parity or spin-triplet even-parity, 
even if the symmetry of the bulk superconductor has a standard even-frequency 
pairing, $i.e$., spin-singlet even-parity or spin-triplet odd-parity~\cite{Berezinskii,tanaka12,Linder2017}, 
due to the mixing of parity~\cite{odd3,Eschrig2007,odd3b,PhysRevLett.116.257001,PhysRevB.96.174509}.
Thus, the ZESABS always accompanies an odd-frequency pairing due to the 
breaking of translational invariance in the superconductor%
~\cite{odd3,Eschrig2007,odd3b}. 
If the symmetry of the bulk superconductor is a spin-singlet $d_{xy}$-wave, 
the resulting ZESABSs have an odd-frequency spin-singlet odd-parity 
(OSO, see Table~\ref{table:symbols}) pairing component~\cite{odd3,odd3b}. 
On the other hand, if the symmetry of the bulk superconductor is 
a spin-triplet $p$-wave, the resulting ZESABSs exhibit an
odd-frequency spin-triplet even-parity (OTE, see Table~\ref{table:symbols}) 
pairing symmetry\cite{odd1,odd3,odd3b}.
\begin{table}[htbp]
   \caption{%
      Possible pairing symmetry and 
      symbols for symmetry of Cooper pair amplitude.
      Pairing symmetries of Cooper pair are classified into 
      ESE (even-frequency spin-singlet even-parity),
      ETO (even-frequency spin-triplet odd-parity),
      OSO (odd-frequency  spin-singlet odd-parity), and
      OTE (odd-frequency  spin-triplet even-parity).
   }%
   \begin{tabular}{cccccc}
      \toprule
      \toprule
      frequency & spin & parity & symbol
      \\
      \midrule
      Even & Singlet  & Even  & ESE
      \\
      Even & Triplet  & Odd  & ETO
      \\
      Odd & Singlet  & Odd  & OSO
      \\
      Odd & Triplet  & Even  & OTE
      \\
      \bottomrule
      \bottomrule
   \end{tabular}%
   \label{table:symbols}
\end{table}
In the latter case, if the superconductor is contacted by a diffusive normal metal (DN), 
the OTE pairing can penetrate into the DN~\cite{odd1}, 
as it has an $s$-wave component, which is robust 
against impurity scattering. 
The resulting local density of states (LDOS) in the DN has a 
zero-energy peak (ZEP)~\cite{Proximityp,Proximityp2,Proximityp3,odd1,Tshape} 
in contrast to the conventional proximity effect, 
where the LDOS has a gap-like structure around zero-energy%
~\cite{Golubov88}. This unusual condition is known as 
an anomalous proximity effect%
~\cite{Proximityp,Proximityp2,Proximityp3,Ikegaya2016}. 
The anomalous proximity effect can occur in a DN/noncentrosymmetric 
superconductor junction when the spin-triplet pair potential 
is dominant~\cite{Annuziata}. 

The anomalous proximity effect, triggered by an odd-frequency spin-triplet 
$s$-wave pairing, shows several interesting physical properties:
i) ZEP of the LDOS in the DN%
~\cite{Proximityp,Proximityp2,Higashitani09,Higashitani}, 
ii) ZEP of the LDOS at rough surface~\cite{Bakurskiy}, 
iii) zero bias conductance peak in quasiparticle transport in the DN/spin-triplet 
$p$-wave superconductor junctions~\cite{Proximityp,Proximityp2,Ikegaya2016}, 
iv) significant enhancement of the Josephson current at low temperatures 
in the DN/spin-triplet $p$-wave superconductor junctions~\cite{Proximityp3} 
v) paramagnetic Meissner response%
~\cite{Meissner,WPSKAHBRS98,TAGK05,Meissner10,SuzukiAsano1,
SuzukiAsano2,SuzukiAsano3,Lutchyn2017},
and vi) anomalous surface impedance~\cite{Asano2011,PhysRevB.98.134508}. 
It has been also shown that the anomalous proximity effect can occur 
in topologically designed hybrid systems based on conventional 
spin-singlet $s$-wave superconductor systems with 
spin-orbit coupling and Zeeman effect%
~\cite{Asano2013,Ikegaya,Ikegaya2016}
and the anomalous proximity effect has been 
studied considering the classification of 
the topological nature of the Hamiltonian~\cite{Ikegaya2018}. 
In addition, there are several studies reporting that odd-frequency pairings 
appear as a Majorana fermion, which is a special type of the ZESABS 
in various topological superconducting systems%
~\cite{Asano2013,Stanev,Ebisu,Sau2015,Ebisu2,Klinovaja2016,Cayao,
BalatskyTI,Keidel,Breunig,Crepin,Kashuba}. 
\par
Although OTE $s$-wave pairing has been discussed in 
diffusive ferromagnet (DF)/conventional spin-singlet $s$-wave superconductor 
junctions, it is difficult to 
realize pure OTE $s$-wave pairing state in a DF%
~\cite{Efetov1,Efetov2,volkov,eschrig_pt,Eschrig2003,Asano2007PRL,Braude,
Yokoyama2007,Asano2007PRB,Eschrig2008,Eschrig2015,PhysRevB.92.014508}. 
Conversely, the demonstrated anomalous proximity effect 
generated by ZESABSs specific to unconventional superconductors 
is remarkable, as it can induce a purely odd-frequency pairing or 
at least a significant amount of it at low energy. 
Thus, it is interesting to understand the mechanism of effect of several 
external perturbations on the anomalous proximity effect. 
Among these, the Rashba spin-orbit coupling is 
very interesting, as it inevitably exists near the 
interface or the thin films of superconductor grown on the substrate. 
Here, we focus on the effect of the Rashba spin-orbit coupling 
(RSOC) on the anomalous proximity effect. 
As shown by a study of noncentrosymmetric superconductors, 
the RSOC can mix spin-triplet odd-parity pairing with the 
spin-singlet even-parity 
pairing~\cite{Bauer,Interface,Gorkov,Frigeri,Fujimoto1,Fujimoto09,
PhysRevB.92.134512,PhysRevB.95.184518,PhysRevB.98.075425}. 
It is possible that the
anomalous proximity effect can be induced even in a
spin-singlet $d_{xy}$-wave superconductor, when an
OTE $s$-wave pairing can be induced from the RSOC\@.

In this paper, we study the anomalous proximity effect 
of DN/superconductor junctions with the RSOC by choosing a spin-triplet $p_x$-wave pairing
with a $\mathbf{d}$-vector parallel to $z$-axis and a spin-singlet $d_{xy}$-wave pairing.
As the RSOC, we consider that it is
proportional to $\mathbf{g}\cdot\boldsymbol{\sigma}$ 
with $\mathbf{g}=(\sin k_y,-\sin k_x,0)$ which is originated from 
the inversion symmetry breaking in the $z$-direction and Pauli matrix $\boldsymbol{\sigma}$
in the spin-space.
We calculate the surface density of states, the spin-triplet $s$-wave component 
of the odd-frequency pairing, and the local density of states 
in the DN attached to superconductors, based on a tight-binding model. 
For the spin-triplet $p_{x}$-wave superconductor, 
as the ZESABS is fragile against 
the RSOC~\cite{Kobayashi2014,Kobayashi2015}, 
the magnitude of the OTE pairing is reduced and the 
resulting anomalous proximity effect is weakened by the RSOC\@.
On the other hand, it is shown
that the anomalous proximity effect can be switched on by the RSOC 
for spin-singlet $d_{xy}$-wave superconductor junctions. 
The magnitude of the LDOS at zero-energy in the DN increases with the 
increase of the RSOC\@.
This indicates that 
the anomalous proximity effect can be detected for
high $T_\mathrm{c}$ cuprate junctions.
\par
The organization of this paper is as follows. 
In Section~\ref{sec:open}, we discuss the momentum resolved surface density 
of states and the OTE pairing for semi-infinite superconductors with the RSOC\@.
In Section~\ref{sec:junction}, we consider the DN/superconductor junctions
and discuss the LDOS and the OTE pairing in the DN\@.
In Section~\ref{sec:summary}, we summarize our results. 


\section{Semi-infinite superconductor\label{sec:open}}
\begin{figure}[htbp]
   \centering
   \includegraphics[width = 5.0cm]{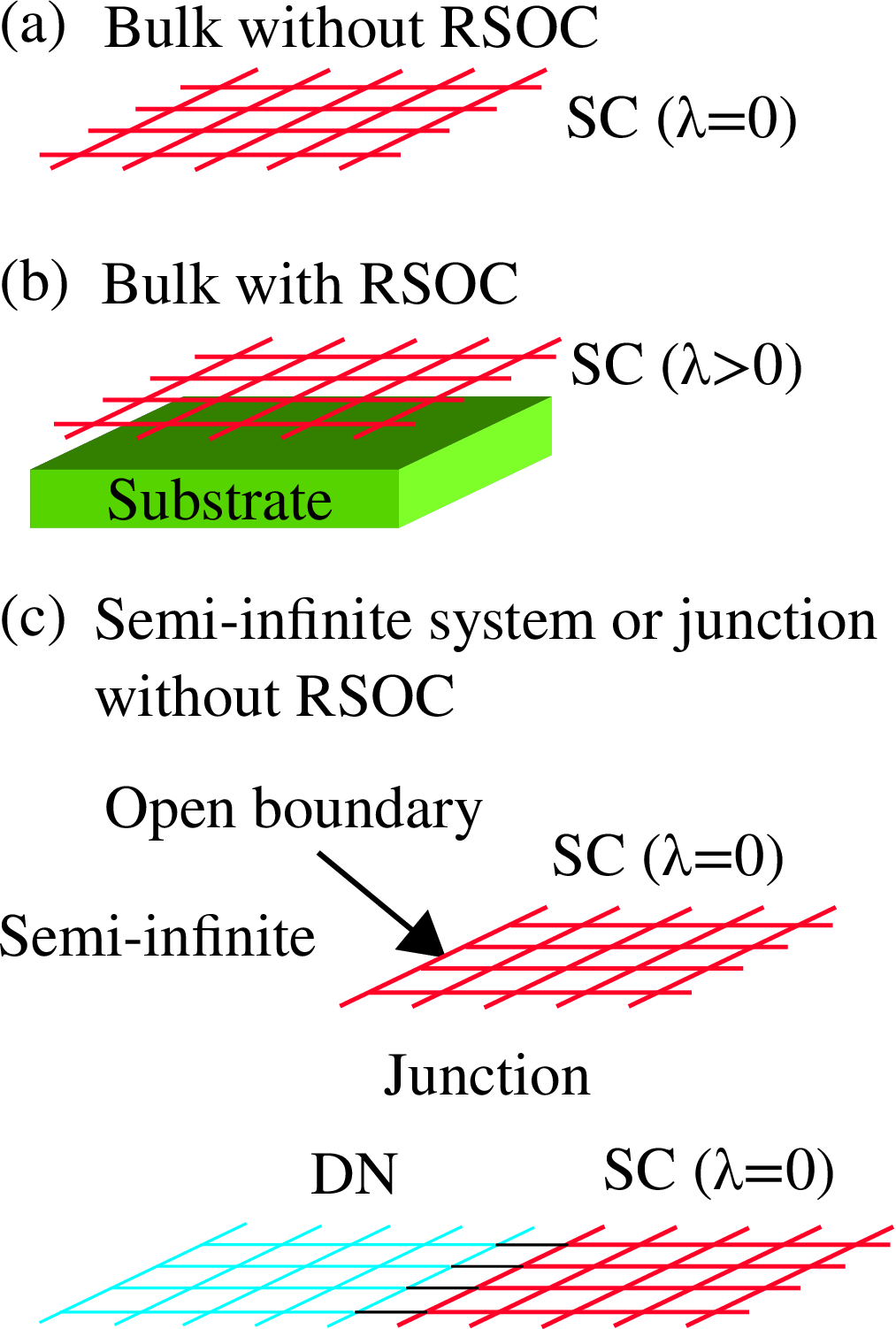}
   \caption{%
      Schematic pictures corresponding to Table~\ref{table:dxy_symmetry}.
      (a) Bulk system without RSOC, 
      (b) Bulk system with RSOC,
      (c) Semi-infinite system or junction without RSOC\@. 
   }%
   \label{fig:pic_symmetry}
\end{figure}
\begin{table}[htbp]
   \caption{%
      Possible symmetry of Cooper pair amplitude for spin-triplet $p_{x}$-wave superconductor
      and spin-singlet $d_{xy}$-wave superconductor 
      in the bulk system.
      TS indicates translational symmetry in the $x$-direction.
      Symbols in the third and fourth columns are given in Table~\ref{table:symbols}.
      (a), (b) and (c) in the first column correspond to those in Fig.~\ref{fig:pic_symmetry}.
   }%
   \begin{tabular}{cccccc}
      \toprule
      \toprule
      &TS & RSOC & spin-triplet & spin-singlet 
      \\
      & & & $p_x$-wave & $d_{xy}$-wave
      \\
      \midrule
      (a) &YES & NO  & ETO       & ESE
      \\
      (b) &YES & YES & ETO, ESE   & ESE, ETO
      \\
      (c) &NO  & NO  & ETO, OTE   & ESE, OSO
      \\
      \bottomrule
      \bottomrule
   \end{tabular}%
   \label{table:dxy_symmetry}
\end{table}
Before we start discussion,
we explain about possible symmetry class of 
Cooper pair by the external symmetry breaking 
for spin-triplet $p_x$-wave and spin-singlet $d_{xy}$-wave superconductors. 
Pairing symmetry of the spin-triplet $p_x$-wave superconductor 
in the bulk belongs to 
even-frequency spin-triplet odd-parity (ETO)  
as shown in Fig.~\ref{fig:pic_symmetry}~(a) and Table~\ref{table:dxy_symmetry}~(a). 
In the presence of the RSOC, since the spin-rotational symmetry and spatial 
parity are broken at the same time, 
even-frequency spin-singlet even-parity (ESE) pairing 
mixes. Then, the resulting pairing symmetry are ESE and ETO 
[Fig.~\ref{fig:pic_symmetry}~(b) and Table~\ref{table:dxy_symmetry}~(b)]~\cite{Frigeri}. 
When only the translational invariance in the $x$-direction is 
broken, the symmetry of pair amplitude are ETO and 
OTE shown in Fig.~\ref{fig:pic_symmetry}~(c) and Table~\ref{table:dxy_symmetry}~(c)~\cite{tanaka12}. 

On the other hand, the spin-singlet $d_{xy}$-wave 
superconductor in the bulk belongs to 
the even-frequency spin-singlet even-parity (ESE) pairing 
as shown in Fig.~\ref{fig:pic_symmetry}~(a) and Table~\ref{table:dxy_symmetry}~(a). 
In the presence of the RSOC, ETO pairing mixes and  
the resulting pairing symmetry are ESE and ETO 
[Fig.~\ref{fig:pic_symmetry}~(b) and Table~\ref{table:dxy_symmetry}~(b)]~\cite{Frigeri}. 
On the other hand, when only the translational invariance is 
broken without the RSOC, the symmetry of pair amplitude are ESE and 
OSO [Fig.~\ref{fig:pic_symmetry}~(c) and Table~\ref{table:dxy_symmetry}~(c)]~\cite{tanaka12}. 

In the case that the system does not have translational symmetry 
in the presence of the RSOC, all pairing symmetries: ETO, ESE, OTE and OSO exist 
both for spin-triplet $p_{x}$-wave and spin-singlet $d_{xy}$-wave 
superconductors.

\subsection{Model\label{sec:open_model_method}}
\begin{figure}[htbp]
   \centering
   \includegraphics[width = 6.7cm]{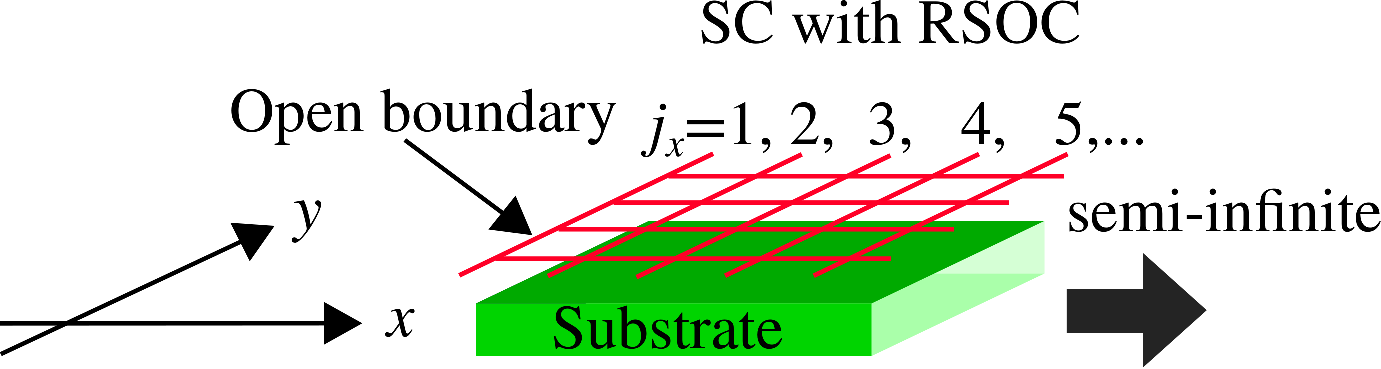}
   \caption{%
      Schematic illustration of semi-infinite lattice model ($j_x\geq1$) SC with RSOC\@.
   }%
   \label{fig:pic_semi}
\end{figure}

We consider two dimensional superconductors on a square lattice model 
with open boundary condition [Fig.~\ref{fig:pic_semi}]
where the superconductor is located in $j_x\geq1$ 
[$\mathbf{j}=(j_x,j_y)$ is the coordinate of a lattice site.]
and discuss the RSOC dependence of a surface density of states and 
the odd-frequency spin-triplet $s$-wave pair amplitude.
We compare the results for the spin-triplet $p_x$-wave superconductor and the spin-singlet $d_{xy}$-wave one.
Throughout this paper, we use lattice constant as a unit of length.
The Hamiltonian ${\cal H}_\mathrm{SC}$ is given by
\begin{align}
   {\cal H}_\mathrm{SC}
   =&
   {\cal H}_t
   +
   {\cal H}_\mathrm{SO}
   +
   {\cal H}_\Delta
   -
   \mu
   \sum_{\mathbf{j}}
   n_{\mathbf{j}},
   \label{eq:Hamiltonian}
   \\
   {\cal H}_t
   =&
   -t
   \sum_{\langle \mathbf{i},\mathbf{j}\rangle,\sigma}
   \left(
      c_{\mathbf{i},\sigma}^\dag c_{\mathbf{j},\sigma}
      +
      \mathrm{H.c.}
   \right),
   \label{eq:H_t}
   \\
   {\cal H}_\mathrm{SO}
   =&
   \frac{\lambda}{2i}
   \sum_{\mathbf{j},\sigma,\sigma'}
   \left(
      c_{\mathbf{j},\sigma}^\dag c_{\mathbf{j}+\mathbf{e}_y,\sigma'}
      -
      c_{\mathbf{j},\sigma}^\dag c_{\mathbf{j}-\mathbf{e}_y,\sigma'}
   \right)
   {%
      \left(
         \sigma_{1}
      \right)
   }_{\sigma,\sigma'}
   \nonumber\\
      &\hspace{-2mm}-
   \frac{\lambda}{2i}
   \sum_{\mathbf{j},\sigma,\sigma'}
   \left(
      c_{\mathbf{j},\sigma}^\dag c_{\mathbf{j}+\mathbf{e}_x,\sigma'}
      -
      c_{\mathbf{j},\sigma}^\dag c_{\mathbf{j}-\mathbf{e}_x,\sigma'}
   \right)
   {%
      \left(
         \sigma_{2}
      \right)
   }_{\sigma,\sigma'},
   \label{eq:HSO_real}
   \\
   {\cal H}_\Delta
   =&
   \sum_{\mathbf{i},\mathbf{j},\sigma,\sigma'}
   \Delta_{\mathbf{i},\mathbf{j}}^{\sigma,\sigma'}
   \left(
      c_{\mathbf{i},\sigma}^\dag c_{\mathbf{j},\sigma'}^\dag
      +
      \mathrm{H.c.}
   \right),
   \label{eq:HSC}
   \\
   n_{\mathbf{j}}
   =&
   \sum_\sigma c_{\mathbf{j},\sigma}^\dag c_{\mathbf{j},\sigma},
\end{align}
where $\langle \mathbf{i},\mathbf{j}\rangle$ in Eq.~(\ref{eq:H_t}) denotes a sum of 
nearest-neighbor pairs,
$\sigma_i$ ($i=1,2,3$) is a Pauli matrix, 
and $\mathbf{e}_x$ ($\mathbf{e}_y$) is a unit vector in the $x$ ($y$)-direction
i.e., $\mathbf{e}_x=(1,0)$ and $\mathbf{e}_y=(0,1)$.
$t$ is a hopping integral, $\lambda$ is a Rashba spin-orbit coupling and
$\mu$ is a chemical potential.
Here we assume $0\leq\lambda$.
We set $\mu/t=-2.4$ where we do not have to take into account 
the effect of the van Hove singularity.
As the symmetry of the superconducting gap, we consider a spin-triplet $p_x$-wave pairing and a spin-singlet $d_{xy}$-wave one.
For the spin-triplet $p_x$-wave case, the pair potential is 
\begin{align}
   \Delta_{\mathbf{i},\mathbf{j}}^{\sigma,\sigma'}
   =&
   \frac{\Delta_0}{i}\delta_{\mathbf{i},\mathbf{j}-\mathbf{e}_x}\delta_{\sigma,-\sigma'},
   \label{eq:Delta_p}
\end{align}
where we choose the direction of the $\mathbf{d}$-vector parallel to the $z$-direction.
For the spin-singlet $d_{xy}$-wave case, the pair potential is 
\begin{align}
   \Delta_{\mathbf{i},\mathbf{j}}^{\sigma,\sigma'}
   =&
   -
   \frac{\Delta_0}{2}
   f(\sigma)
   (
   \delta_{\mathbf{i},\mathbf{j}-\mathbf{e}_x-\mathbf{e}_y}
   -
   \delta_{\mathbf{i},\mathbf{j}-\mathbf{e}_x+\mathbf{e}_y}
   )
   \delta_{\sigma,-\sigma'},
   \label{eq:Delta_d}
\end{align}
with $f(\uparrow)=1$ and $f(\downarrow)=-1$.
We set $\Delta_0/t=0.01$ for both spin-triplet $p_x$-wave and spin-singlet $d_{xy}$-wave superconductor
throughout this paper.

The energy dispersion of the periodic system for the spin-triplet $p_x$-wave superconductor 
is
\begin{align}
   E(\mathbf{k})
   =&
   \pm
   \left|
   \sqrt{\xi_\mathbf{k}^2+{(\Delta_0)}^2\sin^2 k_x}
   \pm
   \lambda
   \sqrt{\sin^2 k_x + \sin^2 k_y}
   \right|,
   \label{eq:dispersion_px}
\end{align}
with $\xi_\mathbf{k} = -2t(\cos k_x+\cos k_y)-\mu$ and the superconducting gap closes 
for $\lambda\geq\Delta_0$.
For the spin-singlet $d_{xy}$-wave superconductor, the energy dispersion 
is
\begin{align}
   &
   E(\mathbf{k})
   \nonumber\\
   =&
   \pm
   \sqrt{%
      {%
         \left[
            \xi_\mathbf{k}
            \pm
            \lambda(\sin k_x+\sin k_y)
         \right]
      }^2
      +
      {%
         \left(
            \Delta_0\sin k_x\sin k_y
         \right)
      }^2
   }.
   \label{eq:dispersion_dxy}
\end{align}

\subsection{Method to calculate local Green's function\label{sec:open_method}}
The surface DOS $\rho(E,k_y)$ and the odd-frequency spin-triplet $s$-wave pair amplitude are 
calculated by using a surface Green's function $G^\infty_{j_x=1}(z,k_y)$.
Here, $G^\infty_{j_x=1}(z,k_y)$ is the Green's function at the leftmost site
[Fig.~\ref{fig:pic_semi}].
The derivation of $G^\infty_{j_x=1}(z,k_y)$ is explained in the Appendix~\ref{sec:App_der_green}.

The momentum resolved surface DOS is obtained as
\begin{align}
   \rho(E,k_y)
   =&
   -
   \frac{1}{\pi}
   \mathrm{Im}
   \left[
      \mathrm{tr}G_{j_x=1}^\infty(z=E+i\eta,k_y)
   \right],
   \label{eq:rho_semi}
\end{align}
where $\eta$ is an infinitesimally small constant 
(we set $\eta/t=10^{-5}$.) and trace is only taken in
particle space.
Then the surface DOS $\bar{\rho}(E)$ is given by
\begin{align}
   \bar{\rho}(E)
   =&
   \frac{1}{2\pi}
   \int_{-\pi}^{\pi} d k_y\rho(E,k_y).
   \label{eq:rho_semi_ave}
\end{align}

The odd-frequency spin-triplet $s$-wave pair amplitude at the surface is defined as
\begin{align}
   P_{S_z=1}^\mathrm{triplet}(\omega_n)
   =&
   P_{\uparrow,\uparrow}(\omega_n),
   \label{eq:pair_amp_S1}
   \\
   P_{S_z=0}^\mathrm{triplet}(\omega_n)
   =&
   \frac{1}{\sqrt{2}}
   \left[
      P_{\uparrow,\downarrow}(\omega_n)+P_{\downarrow,\uparrow}(\omega_n)
   \right],
   \label{eq:pair_amp_S0}
   \\
   P_{S_z=-1}^\mathrm{triplet}(\omega_n)
   =&
   P_{\downarrow,\downarrow}(\omega_n),
   \label{eq:pair_amp_S_1}
   \\
   P^\mathrm{triplet}(\omega_n)
   =&
   \sum_{S_z=-1,0,1}
   \left|
   P_{S_z}^\mathrm{triplet}(\omega_n)
   \right|,
   \label{eq:pair_amp}
\end{align}
with 
\begin{align}
   P_{\sigma,\sigma'}(\omega_n)
   =&
   \frac{1}{N_{k_y}}
   \sum_{k_y}
   \hat{F}_{\sigma,\sigma'}(z=i\omega_n,k_y),
   \\
   G_{j_x=1}^\infty(z,k_y)
   =&
   \begin{pmatrix}
      \hat{G}_{\sigma,\sigma'}(z,k_y) & \hat{F}_{\sigma,\sigma'}(z,k_y)
      \\
      \tilde{F}_{\sigma,\sigma'}(z,k_y) & \tilde{G}_{\sigma,\sigma'}(z,k_y) 
   \end{pmatrix},
   \label{eq:anomalous_G}
\end{align}
where
$\omega_n$ is a Matsubara frequency.
In Eq.~(\ref{eq:anomalous_G}), $\hat{G}_{\sigma,\sigma'}(z,k_y)$,
$\hat{F}_{\sigma,\sigma'}(z,k_y)$, $\tilde{F}_{\sigma,\sigma'}(z,k_y)$
and $\tilde{G}_{\sigma,\sigma'}(z,k_y)$ are $2\times2$ matrices.
$\hat{F}_{\sigma,\sigma'}(z,k_y)$ and $\tilde{F}_{\sigma,\sigma'}(z,k_y)$
are anomalous Green's functions.
Note that $P^\mathrm{triplet}_{S_z}(\omega_n)$ is odd in $\omega_n$.


\subsection{Topological number\label{sec:open_topology}}
In this subsection, we consider a bulk system and explain about a winding 
number $W$ and one-dimensional $\mathbf{Z}_2$ topological number $\nu$. 
$W$ and $\nu$ are topological numbers defined in a bulk
and they predict the number of the ZESABSs
known as bulk-boundary correspondence.
The bulk Hamiltonian for the spin-triplet $p_x$-wave superconductor and
that for the spin-singlet $d_{xy}$-wave superconductor are given by
\begin{align}
   H_p(\mathbf{k})
   =&
   H_t(\mathbf{k})
   +
   H_\mathrm{SO}(\mathbf{k})
   +
   H_\Delta^p(\mathbf{k}),
   \label{eq:Hkp}
   \\
   H_d(\mathbf{k})
   =&
   H_t(\mathbf{k})
   +
   H_\mathrm{SO}(\mathbf{k})
   +
   H_\Delta^d(\mathbf{k}),
\end{align}
with
\begin{align}
   H_t(\mathbf{k})
   =&
   \left[
      -2t
      \left(
         \cos k_x + \cos k_y
      \right)
      -\mu
   \right]
   \sigma_0\tau_3,
   \\
   H_\mathrm{SO}(\mathbf{k})
   =&
   \lambda(\sin k_y\sigma_1\tau_0-\sin k_x\sigma_2\tau_3),
   \label{eq:HSO_k}
   \\
   H_\Delta^p(\mathbf{k})
   =&
   \Delta_0\sin k_x \sigma_1\tau_1,
   \\
   H_\Delta^d(\mathbf{k})
   =&
   -
   \Delta_0\sin k_x \sin k_y \sigma_2\tau_2,
\end{align}
   where $\tau_i$ ($i=1,2,3$) is a Pauli matrix in the particle
   hole space and $\tau_0$ is an identity matrix.
Here $H_t(\mathbf{k})$ and $H_\mathrm{SO}(\mathbf{k})$
are the Fourier transformed form of 
Eq.~(\ref{eq:H_t}) and Eq.~(\ref{eq:HSO_real}), respectively
and $H_\Delta^p(\mathbf{k})$ and $H_\Delta^d(\mathbf{k})$
are the Fourier transformed form of
Eq.~(\ref{eq:HSC}) with Eq.~(\ref{eq:Delta_p}) and Eq.~(\ref{eq:Delta_d}), respectively.

\subsubsection{Spin-triplet $p_x$-wave superconductor}
In this subsection, we explain about the Topological numbers for the spin-triplet
$p_x$-wave superconductor.
The winding number can be defined in a bulk if there is a chiral operator $\Gamma$ 
which anticommutes with the bulk Hamiltonian ($\{H(\mathbf{k}),\Gamma\}=0$).
The winding number is given by
\begin{align}
   W(k_\parallel) 
   =&
   \frac{i}{4\pi}
   \int_{-\pi}^\pi dk_\perp
   \mathrm{Tr}
   \left[
      \Gamma
      H^{-1}(\mathbf{k})
      \partial_{k_\perp}
      H(\mathbf{k})
   \right],
\end{align}
where $k_\parallel=k_y$ and $k_\perp=k_x$ are wave numbers which are parallel and 
perpendicular to the surface, respectively.

Without the RSOC, 
the chiral operator is 
$\Gamma=\Gamma_p\equiv S_{z} CT=i\sigma_1\tau_2$ 
($S_z=\sigma_3\tau_3$ is a spin-rotational operator around the $z$-axis, 
$C=\sigma_0\tau_1K$ is a charge conjugation operator,
$T=i\sigma_2\tau_0K$ is a time-reversal operator and $K$ is a complex conjugation operator.)
and it anticommutes with $H_t(\mathbf{k})$ and $H_\Delta^p(\mathbf{k})$:
\begin{align}
   \{H_t(\mathbf{k}),\Gamma_p\}=\{H_\Delta^p(\mathbf{k}),\Gamma_p\}
   =&0.
\end{align}
On the other hand,
the RSOC term of the Hamiltonian does not anticommutes with $\Gamma_p$:
\begin{align}
   \{H_\mathrm{SO}(\mathbf{k}),\Gamma_p\}\neq&0.
\end{align}
Therefore, the spin-triplet $p_x$-wave superconductor with non-zero value of the RSOC does not have chiral symmetry and
the winding number cannot be defined.
More precisely, the Hamiltonian Eq.~(\ref{eq:Hkp}) anticommutes with
$\sigma_2\tau_1$ but the winding number is always zero regardless of the value of the RSOC
and it is not related to the number of the surface states.
Then
the ZESABSs are fragile against the 
RSOC given by Eq.~(\ref{eq:HSO_k})~\cite{Kobayashi2014,Kobayashi2015}.
This result does not depend on the direction of the $\mathbf{d}$-vector.
The winding number for the spin-triplet $p_x$-wave superconductor without the RSOC is
\begin{equation}
   W(k_y)
   =
   \begin{cases}
      0 & k_y<-\alpha^p_-,
      \\
      -2 & -\alpha^p_-<k_y<\alpha^p_-,
      \\
      0 & \alpha^p_- <k_y,
   \end{cases}
\end{equation}
for $-4t<\mu<0$ and
\begin{equation}
   W(k_y)
   =
   \begin{cases}
      -2 & -\pi<k_y<-\alpha^p_+,
      \\
      0 & -\alpha^p_+<k_y<\alpha^p_+,
      \\
      -2 & \alpha^p_+ <k_y\leq\pi,
   \end{cases}
\end{equation}
for $0<\mu<4t$ 
with $\alpha^p_\pm=\arccos (-\mu/2t\pm1)$.


Another topological number ($\mathbf{Z}_2$ topological number~\cite{PhysRevB.88.134523})
   can be defined for the spin-triplet $p_x$-wave 
   superconductor in the presence of the RSOC for $k_y=0$ or $\pi$.
At $k_y=0$ or $\pi$, the Hamiltonian given by Eq.~(\ref{eq:Hkp}) possesses time-reversal
symmetry ($T=i\sigma_2\tau_0 K$) and particle-hole symmetry ($C=i\sigma_0\tau_1 K$):
\begin{align}
   T^{-1}H_p(k_x,k_y=\eta)T
   &=
   H_p(-k_x,k_y=\eta),
   \\
   C^{-1}H_p(k_x,k_y=\eta)C
   &=
   -
   H_p(-k_x,k_y=\eta),
\end{align}
with $\eta=0$ or $\pi$.
The eigenvectors of this Hamiltonian has a relation called Kramers degeneracy, i.e.,
\begin{align}
   |u_\alpha^I(-k_x,k_y=\eta)\rangle
   =&
   -e^{i\chi_{\alpha,\eta}(k_x)}T
   |u_\alpha^{II}(k_x,k_y=\eta)\rangle,
   \\
   |u_\alpha^{II}(-k_x,k_y=\eta)\rangle
   =&
   e^{i\chi_{\alpha,\eta}(-k_x)}T
   |u_\alpha^{I}(k_x,k_y=\eta)\rangle,
\end{align}
where $I$ and $II$ indicate the Kramers degeneracy index, 
$\alpha$ denotes the band index and $\chi_{\alpha,\eta}(k_x)$ is a $U(1)$ phase factor.
The Hamiltonian given by Eq.~(\ref{eq:Hkp}) with $k_y=0$ or $\pi$
belongs to DIII class in one dimension and it is characterized by $\mathbf{Z}_2$ topological number.
$\mathbf{Z}_2$ topological number is defined by the one of the
Kramers pair as
\begin{align}
   {(-1)}^{\nu^I_\eta}
   =&
   {(-1)}^{\nu^{II}_\eta}
   =
   \pm1,
\end{align}
with
\begin{align}
   \nu^\beta_\eta
   =&
   \exp
   \left[
      i\int_0^{2\pi} dk_x
      A^{\beta}_\eta(k_x)
   \right],
   \\
   A^\beta_{\eta}(k_x)
   =&
   -i\sum_{\alpha:\mathrm{occ}}\langle u^\beta_\alpha (k_x,k_y=\eta)|
   \partial_{k_x}|u^\beta_\alpha(k_x,k_y=\eta)\rangle.
   \label{eq:Berry_connection}
\end{align}
Here $\beta=I$ or $II$, summation in Eq.~(\ref{eq:Berry_connection}) runs over
its occupied bands and
$A^\beta_{\eta}(k_x)$ is a Berry connection.
The system is topologically trivial when ${(-1)}^{\nu^I_\eta}=1$
and is topologically nontrivial when ${(-1)}^{\nu^I_\eta}=-1$.
When ${(-1)}^{\nu^I_\eta}$ is $-1$, there are ZESABSs.
$\mathbf{Z}_2$ topological number for the spin-triplet $p_x$-wave superconductor with the RSOC 
is 
\begin{align}
   {(-1)}^{\nu^I_{\eta=0}}
   =&
   \begin{cases}
      1 & \mu<-4t \:\:\mathrm{or}\:\: 0<\mu,
      \\
      -1 & -4t<\mu<0,
   \end{cases}
\end{align}
for $k_y=0$ and it is
\begin{align}
   {(-1)}^{\nu^I_{\eta=\pi}}
   =&
   \begin{cases}
      1 & \mu<0 \:\:\mathrm{or}\:\: 4t<\mu,
      \\
      -1 & 0<\mu<4t,
   \end{cases}
\end{align}
for $k_y=\pi$.
\subsubsection{Spin-singlet $d_{xy}$-wave superconductor}
For the spin-singlet $d_{xy}$-wave case, 
ZESABSs can be understood by only using the winding number.
The chiral operator is given by $\Gamma=\Gamma_d\equiv CT=-\sigma_2\tau_1$,
and $\Gamma_d$ anticommutes with the Hamiltonian:
\begin{align}
   \{H_t(\mathbf{k}),\Gamma_d\}=\{H_\Delta^d(\mathbf{k}),\Gamma_d\}
   =\{H_\mathrm{SO}(\mathbf{k}),\Gamma_d\}=0.
\end{align}
Then, the ZESABSs for 
the spin-singlet $d_{xy}$-wave superconductor are robust against the RSOC\@.
The winding number for the spin-singlet $d_{xy}$-wave superconductor with the RSOC 
is complicated but
for the parameters we choose in this paper 
($\lambda/\Delta_0\leq10$ with $\Delta_0/t=0.01$ and $\mu/t=-2.4$), 
the winding number is
\begin{equation}
   W(k_y)
   =
   \begin{cases}
      0 & k_y<-\alpha^d_-,
      \\
      -1 & -\alpha^d_-<k_y<-\alpha^d_+,
      \\
      -2 & -\alpha^d_+<k_y<0,
      \\
      2 & 0<k_y<\alpha^d_+,
      \\
      1 & \alpha^d_+<k_y<\alpha^d_-,
      \\
      0 & \alpha^d_- <k_y,
   \end{cases}
   \label{eq:W_dxy}
\end{equation}
with
\begin{align}
   \cos
   \alpha^d_\pm
   =&
   \frac{1}{4t^2+\lambda^2}
   \left[
      -
      2
      t(2t+\mu)
      \pm
      \lambda
      \sqrt{%
         \lambda^2
         -
         4t\mu
         -
         \mu^2
      }%
   \right].
\end{align}
Other cases with different parameters are shown in
the Appendix~\ref{sec:App_topo_dxy}.

\subsection{Results\label{sec:open_results}}
We discuss the momentum resolved surface DOS, the surface DOS
and the odd-frequency spin-triplet $s$-wave pair amplitude at the surface
for the spin-triplet $p_x$-wave and the spin-singlet $d_{xy}$-wave
superconductors.
\subsubsection{Spin-triplet $p_x$-wave superconductor}
\begin{figure*}[htbp]
   \centering
   \includegraphics[width = 17cm]{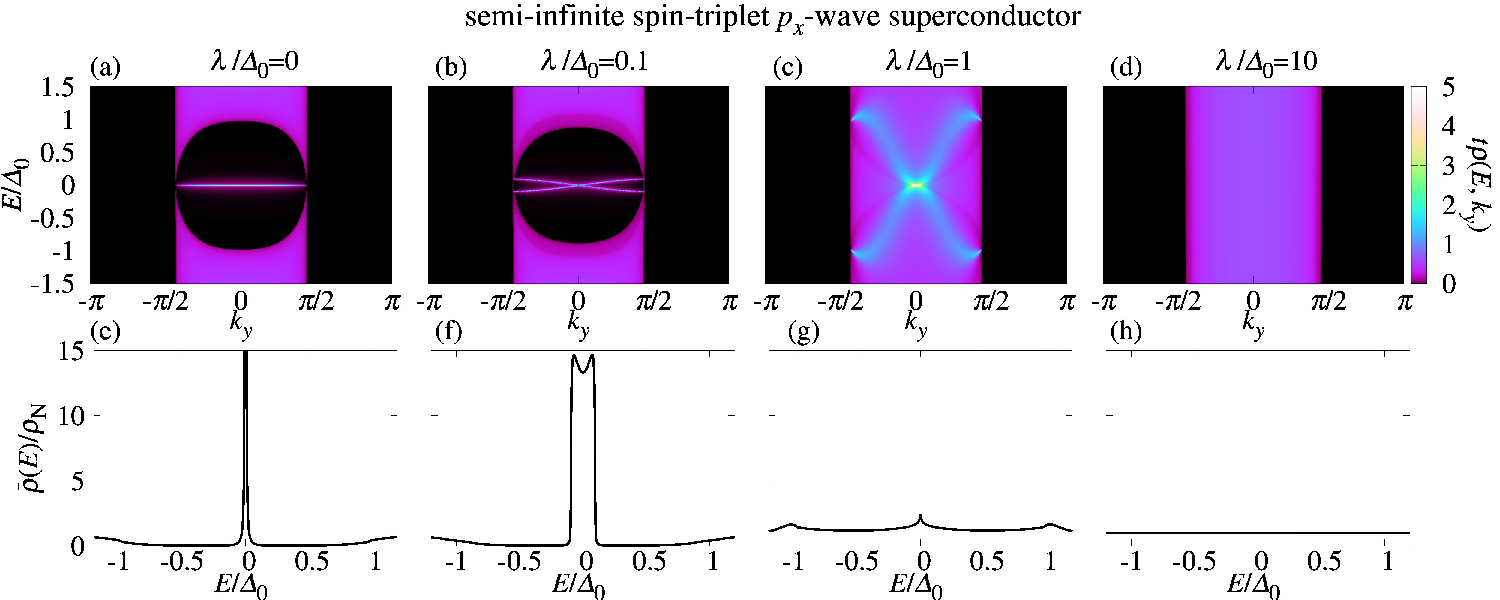}
   \caption{%
      The momentum resolved surface DOS [$\rho(E,k_y)$ from (a) to (d)] 
      and the surface DOS [$\bar{\rho}(E)$ from (e) to (h)]
      for the spin-triplet $p_x$-wave superconductor are plotted for several $\lambda/\Delta_0$.
      Here $\rho_\mathrm{N}$ is a surface density of states at zero-energy with the normal state ($\Delta_0=0$).
      $\rho(E,k_y)$ and $\bar{\rho}(E)$ are calculated at $j_x=1$ (surface)
      in the system shown in Fig.~\ref{fig:pic_semi}.
      The figures in the same column have the same $\lambda$. 
      [(a), (e)] $\lambda=0$, 
      [(b), (f)] $\lambda/\Delta_0=0.1$,
      [(c), (g)] $\lambda/\Delta_0=1$, and
      [(d), (h)] $\lambda/\Delta_0=10$ with $\Delta_0/t=0.01$.
   }%
   \label{fig:LDOS_px_z}
\end{figure*}
\begin{figure}[htbp]
   \centering
   \includegraphics[width = 8.5cm]{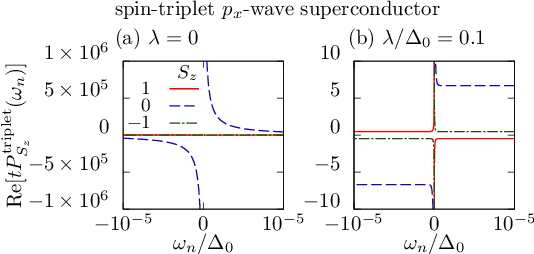}
   \caption{%
      The real part of the odd-frequency spin-triplet $s$-wave pair amplitude 
      $P_{S_z}^\mathrm{triplet}(\omega_n)$ at the surface
      is plotted as a function of $\omega_n/\Delta_0$ for $S_z=1$, $0$, $-1$
      for the spin-triplet $p_x$-wave superconductor.
         (a) $\lambda/\Delta_0=0$
         and 
         (b) $\lambda/\Delta_0=0.1$.
      The imaginary part of $P_{S_z}^\mathrm{triplet}(\omega_n)$ is zero within numerical accuracy.
   }%
   \label{fig:pair_px_ky}
\end{figure}
At first, we discuss the surface DOS  for the spin-triplet $p_x$-wave superconductor.
In Fig.~\ref{fig:LDOS_px_z}, we show the momentum resolved surface DOS $\rho(E,k_y)$ 
given by Eq.~(\ref{eq:rho_semi})
[Figs.~\ref{fig:LDOS_px_z}~(a) to (d)] and the surface DOS $\bar{\rho}(E)$ 
given by Eq.~(\ref{eq:rho_semi_ave})
[Figs.~\ref{fig:LDOS_px_z}~(e) to (h)].
As mentioned in Sect.~\ref{sec:open_topology}, without the RSOC,
zero-energy surface Andreev bound states (ZESABSs) 
appear as a flat band edge state for $-\alpha^p_-<k_y<\alpha^p_-$ 
with $\alpha^p_-=\arccos (-\mu/2t-1)$
[Fig.~\ref{fig:LDOS_px_z}~(a)] due to the non-zero
value of the winding number.
$\bar{\rho}(E)$ also exhibits sharp zero-energy peak (ZEP) as shown in Fig.~\ref{fig:LDOS_px_z}~(e).
Since ZESABSs for the spin-triplet $p_x$-wave superconductor are fragile against 
the RSOC~\cite{Kobayashi2014,Kobayashi2015} 
(see Sect.~\ref{sec:open_topology}),
ZESABSs split into two with the increase of the RSOC other than $k_y=0$
[Figs.~\ref{fig:LDOS_px_z}~(b) and (f)].
As explained in Sect.~\ref{sec:open_topology},
at $k_y=0$, $\mathbf{Z}_2$ topological number has a nontrivial value when $-4t<\mu<0$
and the ZESABS at $k_y=0$ is topologically protected provided that 
the bulk superconducting gap opens 
($\lambda<\Delta_0$).
At $\lambda=\Delta_0$, the bulk gap closes 
[Eq.~(\ref{eq:dispersion_px}) and Figs.~\ref{fig:LDOS_px_z}~(c) and (d)] 
and for $\lambda\gtrsim\Delta$, the corresponding $\bar{\rho}(E)$ is almost independent of $E$ 
[Figs.~\ref{fig:LDOS_px_z}~(g) and (h)].

Next we discuss
the odd-frequency spin-triplet $s$-wave pair amplitude at the surface
[$P_{S_z}^\mathrm{triplet}(\omega_n)$ given in Eqs.~(\ref{eq:pair_amp_S1}) to 
(\ref{eq:pair_amp_S_1})] 
shown in Fig.~\ref{fig:pair_px_ky}.
As can be seen in these graphs, $P_{S_z}^\mathrm{triplet}(\omega_n)$ is odd in $\omega_n$
due to the Fermi-Dirac statistics.
It is also noted that $P_{S_z}^\mathrm{triplet}(\omega_n)$
satisfies $P_{S_z=1}^\mathrm{triplet}(\omega_n)+P_{S_z=-1}^\mathrm{triplet}(\omega_n)=0$ 
within numerical accuracy.
For the spin-triplet $p_{x}$-wave superconductor without the RSOC 
[Fig.~\ref{fig:pair_px_ky}~(a)], $P^\mathrm{triplet}_{S_z=0}(\omega_n)$
has a large value due to the translational symmetry breaking
and $P_{S_z=\pm1}^\mathrm{triplet}(\omega_n)$ is zero due to the spin-rotational symmetry.
$P_{S_z=0}^\mathrm{triplet}(\omega_n)$ drastically 
decreases with the increase of $\lambda$ [Figs.~\ref{fig:pair_px_ky}~(b)]
since the zero-energy flat band disappears other than $k_y=0$. 
On the other hand, divergent behavior very close to $\omega_n=0$
remains due to ZESABS at $k_y=0$.
In Figs.~\ref{fig:pair_ky} (a) to (d) in the Appendix~\ref{sec:App_odd_pair}, 
we also show the odd-frequency $s$-wave pair amplitude at the surface
with $\lambda/\Delta_0=0$, $0.1$, $1$ and $10$.

\begin{figure}[htbp]
   \centering
   \includegraphics[width = 8.0cm]{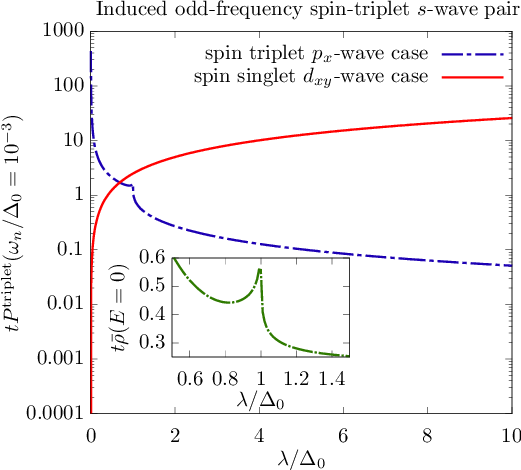}
   \caption{%
      The odd-frequency spin-triplet $s$-wave pair amplitude 
      $P^\mathrm{triplet}(\omega_n/\Delta_0=10^{-3})$
      given by Eq.~(\ref{eq:pair_amp}) is plotted 
      as a function of the RSOC $\lambda$ for 
      the spin-triplet $p_x$-wave superconductor and the spin-singlet $d_{xy}$-wave superconductor.
      In the inset, the surface DOS $\bar{\rho}(E=0)$ is plotted as a function of $\lambda$.
      Here $\Delta_0/t=0.01$.
   }%
   \label{fig:odd_freq_semi}
\end{figure}

The total odd-frequency spin-triplet $s$-wave pair amplitude at the surface
$P^\mathrm{triplet}(\omega_n)$ given by Eq.~(\ref{eq:pair_amp}) with 
$\omega_n/\Delta_0=10^{-3}$ is shown in Fig.~\ref{fig:odd_freq_semi}.
For the spin-triplet $p_x$-wave case, $P^\mathrm{triplet}(\omega_n)$
has a large value at $\lambda=0$.
On the other hand, with the non-zero value of the RSOC, 
$P^\mathrm{triplet}(\omega_n)$ decreases drastically 
due to the absence of the ZESABS other than $k_y=0$.
$P^\mathrm{triplet}(\omega_n)$
has a peak approximately at $\lambda/\Delta_0=1$ where 
the bulk energy gap closes.
The surface DOS at zero-energy also exhibits a similar behavior 
[In the inset of Fig.~\ref{fig:odd_freq_semi}] since the slope 
of the dispersive surface Andreev bound state at $k_y=0$ becomes smaller as 
$\lambda/\Delta_0=1-\delta$ 
approaches unity from $\delta\rightarrow+0$.

\subsubsection{Spin-singlet $d_{xy}$-wave superconductor}
\begin{figure*}[htbp]
   \centering
   \includegraphics[width = 17cm]{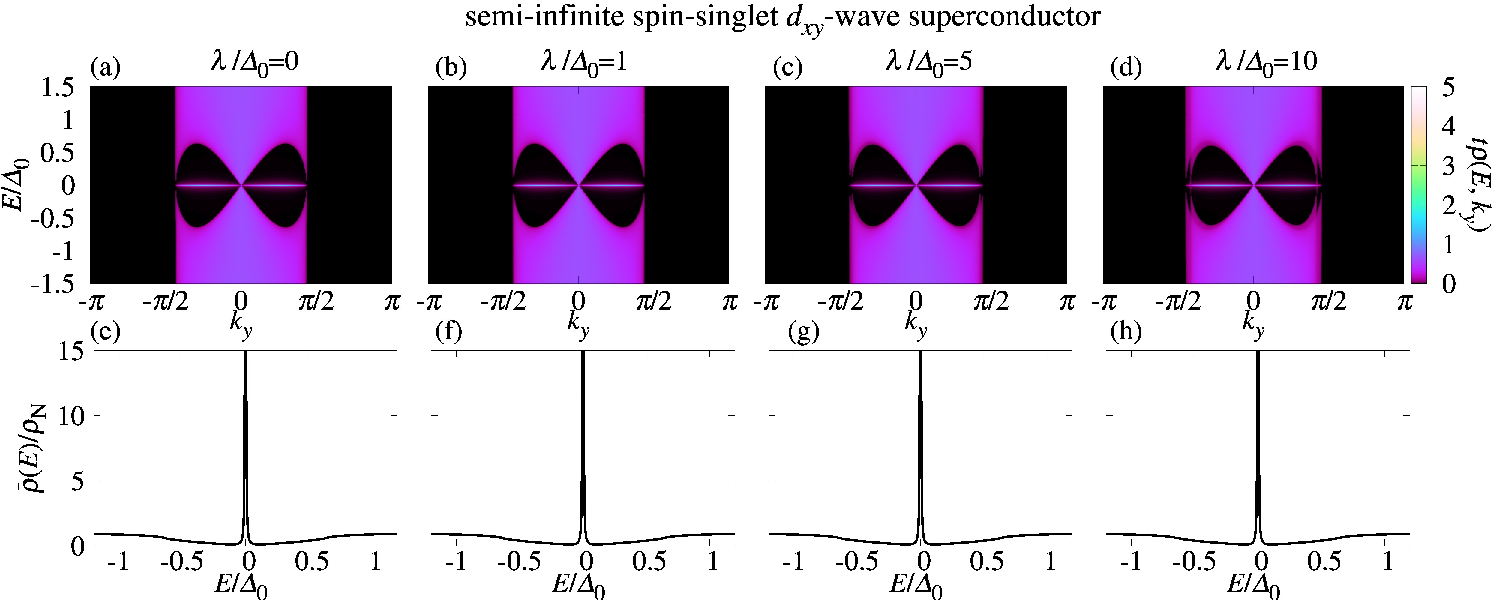}
   \caption{%
      The momentum resolved surface DOS [$\rho(E,k_y)$ from (a) to (d)] 
      and the surface DOS [$\bar{\rho}(E)$ from (e) to (h)]
      for the spin-singlet $d_{xy}$-wave superconductor are plotted for several $\lambda/\Delta_0$.
      Here $\rho_\mathrm{N}$ is a surface density of states at zero-energy with the normal state ($\Delta_0=0$).
      $\rho(E,k_y)$ and $\bar{\rho}(E)$ are calculated at $j_x=1$ (surface)
      in the system shown in Fig.~\ref{fig:pic_semi}.
      The figures in the same column have the same $\lambda$. 
      [(a), (e)] $\lambda=0$, 
      [(b), (f)] $\lambda/\Delta_0=1$,
      [(c), (g)] $\lambda/\Delta_0=5$, and
      [(d), (h)] $\lambda/\Delta_0=10$ with $\Delta_0/t=0.01$.
   }%
   \label{fig:LDOS_dxy}
\end{figure*}
In Fig.~\ref{fig:LDOS_dxy}, $\rho(E,k_y)$ and $\bar{\rho}(E)$ for 
the spin-singlet $d_{xy}$-wave superconductor are shown.
As explained in Sect.~\ref{sec:open_topology}, ZESABSs for 
the spin-singlet $d_{xy}$-wave superconductor are robust against 
the RSOC shown in Figs.~\ref{fig:LDOS_dxy}~(a) to (d).
The regime for the ZESABSs are given by Eq.~(\ref{eq:W_dxy}).
From Eq.~(\ref{eq:dispersion_dxy}), for a 
non-zero value of $\lambda$, 
degeneracy of the point nodes is lifted.
The splitting of the point nodes is too small to recognize in 
Figs.~\ref{fig:LDOS_dxy}~(b) and (c),
but for $\lambda/\Delta_0=10$, it can be seen in Fig.~\ref{fig:LDOS_dxy}~(d).
The surface DOS is almost independent of the strength of the RSOC 
as shown in Figs.~\ref{fig:LDOS_dxy}~(e) to (h) due to the existence of the ZESABSs.

The odd-frequency spin-triplet $s$-wave pair amplitude is absent for 
$\lambda=0$ due to the spin-rotational symmetry.
It is noted that
the odd-frequency spin-singlet $p_y$-wave pair amplitude has 
a large value [Fig.~\ref{fig:omega_dep_dxy_py}~(a) in the Appendix~\ref{sec:App_dxy_py}].
On the other hand, $P_{S_z}^\mathrm{triplet}(\omega_n)$ increases with the increase of 
$\lambda$ [Fig.~\ref{fig:pair_dxy_ky} and Figs.~\ref{fig:pair_ky}~(e) to (f) in the Appendix~\ref{sec:App_odd_pair}].
The amplitude of the $P_{S_z=\pm1}^\mathrm{triplet}(\omega_n)$ 
is larger than that of the $P_{S_z=0}^\mathrm{triplet}(\omega_n)$ for $\lambda>0$ 
since the $S_z=0$ spin-triplet component
of the anomalous Green's function in the bulk system is zero
[Appendix~\ref{sec:App_Green_dxy}].

The total odd-frequency spin-triplet $s$-wave pair amplitude at the surface
$P^\mathrm{triplet}(\omega_n)$ given by Eq.~(\ref{eq:pair_amp}).
$P^\mathrm{triplet}(\omega_n)$ is zero at $\lambda=0$ due to the spin-rotational symmetry 
and it becomes larger with the increase of $\lambda$.
From this result, we can expect that the anomalous proximity effect occurs for 
the spin-singlet $d_{xy}$-wave case with a large value of the RSOC\@.
\begin{figure}[htbp]
   \centering
   \includegraphics[width = 6.2cm]{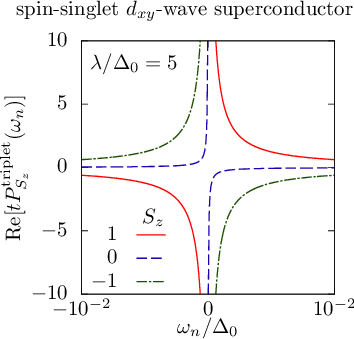}
   \caption{%
      The real part of the odd-frequency spin-triplet $s$-wave pair amplitude 
      $P_{S_z}^\mathrm{triplet}(\omega_n)$ at the surface
      is plotted as a function of $\omega_n/\Delta_0$ for $S_z=1$, $0$, $-1$
      for the spin-singlet $d_{xy}$-wave superconductor with $\lambda/\Delta_0=5$.
      The imaginary part of $P_{S_z}^\mathrm{triplet}(\omega_n)$ is zero within numerical accuracy.
   }%
   \label{fig:pair_dxy_ky}
\end{figure}

\section{Diffusive normal metal/Superconductor junction\label{sec:junction}}
\begin{figure}[htbp]
   \centering
   \includegraphics[width = 8.5cm]{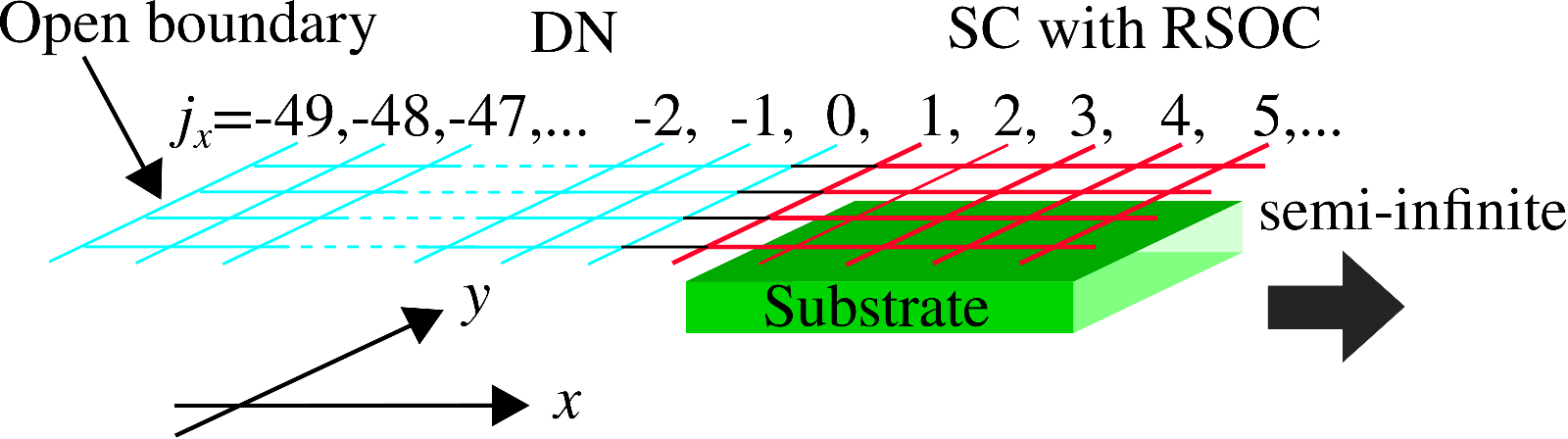}
   \caption{%
      Schematic illustration of 
      DN without RSOC ($-49\leq j_x\leq0$)/semi-infinite ($j_x\geq1$) SC with RSOC junction.
   }%
   \label{fig:pic_junction}
\end{figure}

In this section, we discuss the local DOS and 
the odd-frequency spin-triplet $s$-wave pair amplitude 
in the diffusive normal metal (DN) for DN/SC junctions with the RSOC 
(shown in Fig.~\ref{fig:pic_junction}).
\subsection{Model and Method\label{sec:Model_junction}}
The Hamiltonian for the SC ($j_x\geq1$) is given by Eq.~(\ref{eq:Hamiltonian}) 
and the Hamiltonian for the DN ($-49\leq j_x\leq0$) is
\begin{align}
   &{\cal H}_{\mathrm{DN}}^{l_s}
   \nonumber\\
   &=
   -t\sum_{\langle \mathbf{i},\mathbf{j}\rangle,\sigma}
   \left(
      c_{\mathbf{i},\sigma}^\dag c_{\mathbf{j},\sigma}
      +
      \mathrm{H.c.}
   \right)
   +
   \sum_{-49\leq j_x\leq0,j_y}
   \left(
      -\mu+V_{\mathbf{j}}^{l_s}
   \right)
   n_{\mathbf{j}},
\end{align}
where 
the summations in the $x$-direction 
run over from $-49$ to $0$ ($-49\leq i_x,j_x\leq 0$),
and those in the $y$-direction run over from 1 to $L_y$
($1\leq i_y,j_y\leq L_y$).
$V_\mathbf{j}^{l_s}$ 
($-t\leq V_{\mathbf{j}}^{l_s} \leq t$ with $l_s=1,2,\ldots,N_\mathrm{sample}$) 
is a randomly chosen onsite impurity potential in the DN and there is no RSOC in the DN\@.
The Hamiltonian which connects the DN and the SC is
\begin{align}
   {\cal H}_\mathrm{connect}
   =&
   -t\sum_{j_y,\sigma}
   \left(
      c_{(0,j_y),\sigma}^\dag c_{(1,j_y),\sigma}
      +
      \mathrm{H.c.}
   \right).
\end{align}

The local Green's function is obtained from the same procedure explained in 
Sect.~\ref{sec:open_method}.
The LDOS in the DN is given by
\begin{align}
   \bar{\rho}(E,j_x)
   &=
   -
   \frac{1}{2N_{k_y} L_y N_\mathrm{sample}}
   \nonumber\\
   &
   \times
   \sum_{j_y=j'_y,k_y,l_s}
   \frac{1}{\pi}
   \mathrm{Im}
   \left[
      \mathrm{tr}
      \tilde{g}_{j_x}^{\mathrm{DN},l_s}(z=E+i\eta,j_y,j'_y,k_y)
   \right],
   \label{eq:rho_DN}
\end{align}
where we use a super unit cell $k_y=\pi m/(N_{k_y}L_{y})$ 
with $m=-N_{k_y}+1,\ldots,N_{k_y}$, $N_{k_y}=10$ 
   and $L_y$ is a length of the super unit cell in the $y$-direction
      used in numerical calculation.
Here, trace is only taken in particle space
and we set $\eta/t=10^{-5}$.
The Green's function in the DN $\tilde{g}_{j_x}^{\mathrm{DN},l_s}(z,j_y,j'_y,k_y)$ in Eq.~(\ref{eq:rho_DN}) 
is given in the Appendix~\ref{sec:App_H_DN_SC} [Eq.~(\ref{eq:tilde_g})].
To calculate the LDOS for the spin-triplet $p_x$-wave superconductor junction, 
we take $L_y=120$ which is the length of the system in the $y$-direction 
and to reduce the system size effect, 60 samples for $V_\mathbf{j}^{l_s}$ 
($l_s=1,2,\ldots,60$, i.e., $N_\mathrm{sample}=60$) are used to 
calculate the averaged value of the LDOS and the odd-frequency pair amplitude
in the DN\@.
We take $L_y=240$ and 12 samples ($N_\mathrm{sample}=12$) are used 
to the impurity average for the spin-singlet $d_{xy}$-wave junction
since size effect for the spin-singlet $d_{xy}$-wave case is larger than 
that for the spin-triplet $p_x$-wave case since the pair potential for
the spin-singlet $d_{xy}$-wave superconductor depends on $k_y$ but that
for the spin-triplet $p_{x}$-wave superconductor does not depend on $k_y$.

To discuss the relation between the anomalous proximity effect and 
the odd-frequency spin-triplet $s$-wave pair amplitude, 
we calculate its averaged value in the DN region 
given by
\begin{align}
   \bar{P}_{S_z=1}^\mathrm{DN,triplet}(\omega_n)
   =&
   \frac{1}{N}
   \sum_{j_x=N_\mathrm{i}}^{N_\mathrm{f}}
   P_{j_x,\uparrow,\uparrow}(\omega_n),
   \label{eq:pair_amp_S1_DN}
   \\
   \bar{P}_{S_z=0}^\mathrm{DN,triplet}(\omega_n)
   =&
   \frac{1}{\sqrt{2}N}
   \sum_{j_x=N_\mathrm{i}}^{N_\mathrm{f}}
   \left[
      P_{j_x,\uparrow,\downarrow}(\omega_n)+P_{j_x,\downarrow,\uparrow}(\omega_n)
   \right],
   \label{eq:pair_amp_S0_DN}
   \\
   \bar{P}_{S_z=-1}^\mathrm{DN,triplet}(\omega_n)
   =&
   \frac{1}{N}
   \sum_{j_x=N_\mathrm{i}}^{N_\mathrm{f}}
   P_{j_x,\downarrow,\downarrow}(\omega_n),
   \label{eq:pair_amp_S_1_DN}
\end{align}
with 
\begin{align}
   &P_{j_x,\sigma,\sigma'}(\omega_n)
   \nonumber
   \\
   =&
   \frac{1}{2N_{k_y} L_y N_\mathrm{sample}}
   \sum_{k_y,j_y=j'_y,l_s}
      \hat{f}_{j_x,\sigma,\sigma'}^{l_s}(i\omega_n,j_y,j'_y,k_y),
   \label{eq:def_p}
\end{align}
where we average the odd-frequency pair amplitude in the DN\@:
$N_\mathrm{i}=-49$, $N_\mathrm{f}=0$ and $N=N_\mathrm{f}-N_\mathrm{i}+1=50$.
Here $2L_y\times2L_y$ matrix $\hat{f}_{j_x,\sigma,\sigma'}^{l_s}(z=i\omega_n,j_y,j'_y,k_y)$
is an anomalous Green's function of $\tilde{g}_{j_x}^{\mathrm{DN},l_s}(z,j_y,j'_y,k_y)$ 
in Eq.~(\ref{eq:rho_DN})
defined as the same manner as in Eq.~(\ref{eq:anomalous_G}).
To calculate the odd-frequency spin-triplet $s$-wave pair amplitude, 
we take the same system size in the $y$-direction and the same number of the samples
are used to the impurity average.
\subsection{Results\label{sec:junction_results}}
We discuss the LDOS and the odd-frequency spin-triplet $s$-wave
pair amplitude in the DN for the spin-triplet $p_x$-wave and
the spin-singlet $d_{xy}$-wave superconductor junction.
\subsubsection{Spin-triplet $p_x$-wave superconductor junction}
\begin{figure*}[htbp]
   \centering
   \includegraphics[width = 17cm]{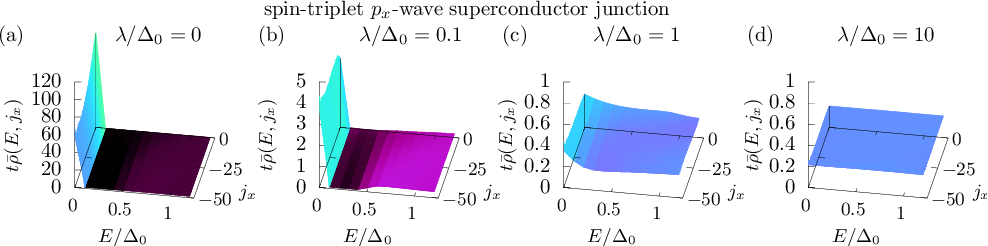}
   \caption{%
      The LDOS [$\bar{\rho}(E,j_x)$] in the DN is plotted as functions 
      of $j_x$ and $E/\Delta_0$ for several strength of the RSOC
      for the spin-triplet $p_x$-wave superconductor junction.
      The amplitude of the RSOC is
      (a) $\lambda/\Delta_0=0$, 
      (b) $0.1$,
      (c) $1$, and
      (d) $10$ with $\Delta_0/t=0.01$.
   }%
   \label{fig:LDOS_DN_px}
\end{figure*}
\begin{figure*}[htbp]
   \centering
   \includegraphics[width = 17cm]{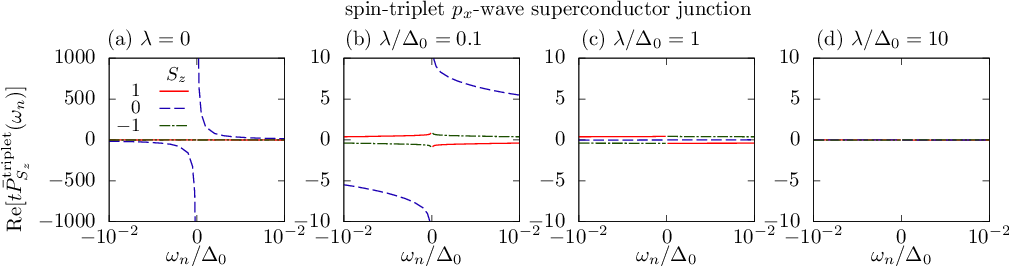}
   \caption{%
      The averaged value of the real part of the odd-frequency spin-triplet $s$-wave pair amplitude
      in the DN $\bar{P}^\mathrm{triplet}_{S_z}(\omega_n)$ is plotted 
      as a function of $\omega_n$ for the spin-triplet $p_x$-wave
      superconductor junction with $S_z=1$, $0$ and $-1$.
      The imaginary part of $\bar{P}^\mathrm{triplet}_{S_z}(\omega_n)$ is zero 
      within numerical accuracy.
      (a) $\lambda/\Delta_0=0$,
      (b) $0.1$,
      (c) $1$, and
      (d) $10$.
   }%
   \label{fig:omega_pair_amp_DN_px_z}
\end{figure*}
Firstly, we discuss the LDOS in the DN 
corresponding to Fig.~\ref{fig:pic_junction}.
LDOS for the spin-triplet $p_x$-wave superconductor junction without the RSOC 
has a sharp ZEP 
as shown in Fig.~\ref{fig:LDOS_DN_px} (a) consistent with the 
previous results in quasiclassical regime~\cite{Proximityp}. 
It is called anomalous proximity effect%
~\cite{Proximityp,Proximityp2,Proximityp3}. 
The ZEP on LDOS in the DN stems from ZESABSs as a flat band zero-energy state
in semi-infinite spin-triplet $p_{x}$-wave superconductor.  
The present ZESABS accompanies OTE $s$-wave pairing, then it can 
penetrate into the DN\@. 
However, after switching on $\lambda$ in the spin-triplet $p_x$-wave superconductor, 
the ZEP of the LDOS in the DN is suppressed by $\lambda$, as shown in 
Fig.~\ref{fig:LDOS_DN_px} (b).
In this case, if we look at the semi-infinite spin-triplet 
$p_{x}$-wave superconductor without the DN,  
the zero-energy flat bands at 
the surface split into two [Figs.~\ref{fig:LDOS_px_z} (b) and (f)] 
and the surface DOS at zero-energy is suppressed. 
This is the reason why the LDOS at zero-energy in the DN is suppressed 
in the DN/spin-triplet $p_x$-wave superconductor junction 
in the presence of the RSOC\@.
For $\lambda$ larger than $\Delta_0$, the bulk energy gap closes [see Figs.~\ref{fig:LDOS_px_z} (c) and (d)]
and the ZEP almost vanishes [Figs.~\ref{fig:LDOS_DN_px} (c) and (d)].

Corresponding to the suppression of the height of ZEP in the DN, 
the averaged value of
the odd-frequency spin-triplet $s$-wave pair amplitude 
$\bar{P}^\mathrm{DN,triplet}_{S_z=0}(\omega_n)$ given by Eq.~(\ref{eq:pair_amp_S0_DN})
in the DN  
becomes smaller with the increase of the RSOC 
[Figs.~\ref{fig:omega_pair_amp_DN_px_z} (a) and (b)]
and this behavior of $\bar{P}^\mathrm{DN,triplet}_{S_z=0}(\omega_n)$ is similar to 
that for the semi-infinite system [Figs.~\ref{fig:pair_px_ky} (a) and (b)].
The odd-frequency spin-triplet $s$-wave pair amplitude $\bar{P}^\mathrm{DN,triplet}_{S_z=\pm1}(\omega_n)$ is zero 
for $\lambda=0$ due to the spin-rotational symmetry and it has 
non-zero value for $\lambda>0$
[Figs.~\ref{fig:omega_pair_amp_DN_px_z} (b) and (c)] and it also satisfies
$\bar{P}^\mathrm{DN,triplet}_{S_z=1}(\omega_n)+\bar{P}^\mathrm{DN,triplet}_{S_z=-1}(\omega_n)=0$
within numerical accuracy.
For large magnitude of $\lambda$, all kinds of $\bar{P}^\mathrm{DN,triplet}_{S_z}(\omega_n)$
becomes almost zero [Fig.~\ref{fig:omega_pair_amp_DN_px_z} (d)].
This result is similar to that for semi-infinite system shown in 
Fig.~\ref{fig:pair_px_ky}
and Figs.~\ref{fig:pair_ky}~(a) to (d)
in the Appendix~\ref{sec:App_odd_pair}.

\subsubsection{Spin-singlet $d_{xy}$-wave superconductor junction}
\begin{figure*}[htbp]
   \centering
   \includegraphics[width = 17cm]{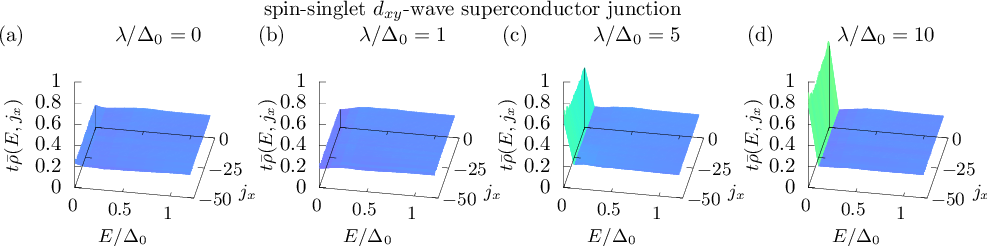}
   \caption{%
      The LDOS [$\bar{\rho}(E,j_x)$] in the DN is plotted as functions 
      of $j_x$ and $E/\Delta_0$ for several strength of the RSOC for 
      the spin-singlet $d_{xy}$-wave superconductor junction.
      The amplitude of the RSOC is
      (a) $\lambda/\Delta_0=0$, 
      (b) $1$,
      (c) $5$, and
      (d) $10$ with $\Delta_0/t=0.01$.
   }%
   \label{fig:LDOS_DN_dxy}
\end{figure*}
\begin{figure}[htbp]
   \centering
   \includegraphics[width = 8.5cm]{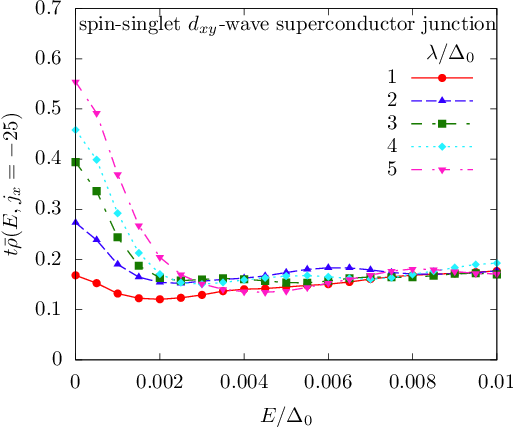}
   \caption{%
      The LDOS for the spin-singlet $d_{xy}$-wave superconductor junction 
      at $j_x=-25$ (center of the DN) is plotted as a function of $E/\Delta_0$ for 
      several $\lambda/\Delta_0$ with $\Delta_0/t=0.01$.
   }%
   \label{fig:ZEP_near0_junction}
\end{figure}

\begin{figure*}[htbp]
   \centering
   \includegraphics[width = 17cm]{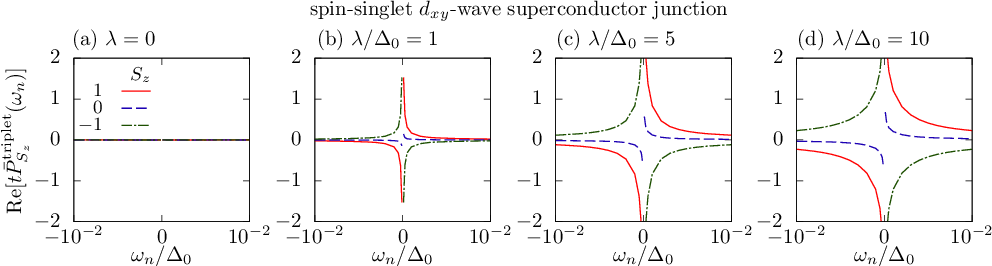}
   \caption{%
      The averaged value of the real part of the odd-frequency spin-triplet $s$-wave pair amplitude
      in the DN $\bar{P}^\mathrm{triplet}_{S_z}(\omega_n)$ is plotted as a function 
      of $\omega_n$ for the spin-singlet $d_{xy}$-wave
      superconductor junction with $S_z=1$, $0$ and $-1$.
      The imaginary part of $\bar{P}^\mathrm{triplet}_{S_z}(\omega_n)$ is zero 
      within numerical accuracy.
      (a) $\lambda/\Delta_0=0$,
      (b) $1$,
      (c) $5$, and
      (d) $10$.
   }%
   \label{fig:omega_pair_amp_DN_dxy}
\end{figure*}
For the spin-singlet $d_{xy}$-wave superconductor 
without the RSOC,
the LDOS in the DN is nearly constant as a function of $E$ and $j_x$ 
[Fig.~\ref{fig:LDOS_DN_dxy} (a)]. 
It is known from the quasiclassical 
theory of proximity effect in unconventional 
superconductor, that
the LDOS is reduced to that in the normal state due to the absence of 
the proximity effect in the DN~\cite{Proximityd,Proximityd2}.
The slight energy dependence is due to the finite size effect in the 
numerical calculation of the lattice model. 
A similar behavior also appears for $\lambda/\Delta_{0}=1$ [Fig.~\ref{fig:LDOS_DN_dxy} (b)].
However, for a large RSOC values ($\lambda/\Delta_0=5$ and $10$), 
the LDOS has a ZEP in the DN region
[Figs.~\ref{fig:LDOS_DN_dxy} (c) and (d)]. 
In Fig.~\ref{fig:ZEP_near0_junction}, we show the LDOS at the center of the DN ($j_x=-25$)
and for $\lambda/\Delta_0\gtrsim2$, the ZEP appears.
Therefore, it may be possible to detect the anomalous proximity effect for 
the spin-singlet $d$-wave superconductor if the amplitude of the RSOC becomes larger than
that of the pair potential.
System size and the number of samples dependence of the LDOS are discussed in the 
Appendix~\ref{sec:App_size}.

At $\lambda=0$, the odd-frequency spin-triplet $s$-wave pair amplitude 
$\bar{P}^{\mathrm{DN,triplet}}_{S_z}(\omega_n)$ 
is zero due to the spin-rotational symmetry [Figs.~\ref{fig:omega_pair_amp_DN_dxy}~(a)].
The amplitude of $\bar{P}^{\mathrm{DN,triplet}}_{S_z}(\omega_n)$ 
increases as the increase of $\lambda$ as shown in 
Figs.~\ref{fig:omega_pair_amp_DN_dxy}~(b) to (d).
Here, non-zero value of the odd-frequency spin-triplet $s$-wave pair amplitude
in the DN means that it is robust against the impurity scattering in the DN
[spatial dependence is shown in Fig.~\ref{fig:DN_SC_onsite} in the
Appendix~\ref{sec:App_spatial_dep_odd}.].
Qualitative behavior of the odd-frequency spin-triplet $s$-wave pair amplitude 
in the DN as a function of the RSOC [Figs.~\ref{fig:omega_pair_amp_DN_dxy}~(a) to (d)]
is the same as that for the semi-infinite SC system at the surface 
i.e., it increases as the increase of the RSOC
[see also Figs.~\ref{fig:pair_ky}~(e) to (h) in the Appendix~\ref{sec:App_odd_pair}].
The odd-frequency spin-triplet $s$-wave pair amplitude also satisfies
$\bar{P}^\mathrm{DN,triplet}_{S_z=1}(\omega_n)+\bar{P}^\mathrm{DN,triplet}_{S_z=-1}(\omega_n)=0$
within numerical accuracy.
The odd-frequency spin-singlet $p_y$-wave pair amplitude rapidly decays 
in the DN since it is fragile against 
impurity scattering [Appendix~\ref{sec:App_spatial_dep_odd}].

\section{summary\label{sec:summary}}
In this paper, we have studied the proximity effect in the DN/superconductor junctions, 
focusing on the anomalous proximity effect, where the LDOS in the DN has a 
zero-energy peak. 
For the spin-triplet $p_{x}$-wave superconductor case, when the direction of 
the $\mathbf{d}$-vector is parallel to the $z$-direction, 
ZESABSs are suppressed by the RSOC $\lambda$ and the anomalous proximity effect 
disappears for large magnitude of $\lambda$.

It should be also noted that 
the anomalous proximity effect is switched on by the 
RSOC in spin-singlet $d_{xy}$-wave superconductor junctions. 
The resulting zero-energy LDOS and the magnitude of the
odd-frequency spin-triplet $s$-wave pair amplitude is 
enhanced with the increase of the magnitude of $\lambda$. 
This indicates that high $T_\mathrm{c}$ cuprate junctions can be used to 
detect the anomalous proximity effect. 
Another candidate material is  CeCoIn$_5$, where the promising 
pairing symmetry is spin-singlet $d$-wave~\cite{PhysRevLett.94.107005}. 
It has been reported that  the amplitude of the RSOC exceeds 
that of the superconducting gap~\cite{PhysRevLett.112.156404} in the 
superlattices of CeCoIn$_5$.

We comment on the effect of surface roughness on the odd-frequency 
spin-triplet  $s$-wave pair amplitude for the spin-singlet $d_{xy}$-wave superconductor.  
It is known that in the presence of SABSs, pair potential decays near the 
surface~\cite{Nagato95,Buchholtz1995,Barash97,Tanuma01}. 
It has been clarified that the surface roughness influences on the 
SABS and the resulting height of ZEP is suppressed by roughness%
\cite{Nagai96,Golubov1999} 
since the induced odd-frequency pairing near the interface 
has a odd-parity with $p_{y}$-wave like symmetry~\cite{LuBo2016}. 
On the other hand, in the presence of the RSOC, we can generate odd-frequency 
spin-triplet $s$-wave pair amplitude near the interface. 
The present pair amplitude is robust against the 
surface diffusive scattering~\cite{Bakurskiy}. 
In order to clarify this point, we have calculated 
spin-triplet $s$-wave pair amplitude in 
diffusive layer attached to spin-singlet $d_{xy}$-wave superconductor 
junction with the RSOC in the presence of 
the spatial depletion of the pair potential near the surface 
[Appendix~\ref{sec:App_tanh}]. 
We have checked the robustness of the odd-frequency spin-triplet $s$-wave
pair amplitude against diffusive scattering.
Even if the spatial depletion of the pair potential near the 
surface exists, the magnitude of the 
odd-frequency spin-triplet $s$-wave pair 
amplitude does not almost change as compared to the case with the 
constant pair potential up to the surface. 

\par

In this paper, we assume that a Rashba spin-orbit coupling 
exists in the two-dimensional superconductor uniformly along the 
$z$-direction, 
as superconductor is assumed to be thin. 
In addition to this effect, spin-orbit coupling also exists near the interface, on the plane 
parallel to the interface~\cite{Samokhin1,Samokhin2}. 
In this case, the direction of the Rashba vector is different. 
For spin-singlet pair potentials, the direction of the spin-orbit coupling
only affects the direction of the spin of the induced odd-frequency spin-triplet
pair amplitude due to the spin-rotational symmetry.
On the other hand,
for spin-triplet pair potentials, the relative angle of the $\mathbf{d}$-vector
and the Rashba spin-orbit coupling affects surface Andreev bound states
like diffusive ferromagnet/superconductor junctions~\cite{PhysRevB.92.024510}.
It is also interesting to study this type of spin-orbit coupling. 
\par

We comment on the case that the amplitude of the RSOC is much larger than the hopping integral
for the spin-singlet $d_{xy}$-wave superconductor.
In our study, we consider that the amplitude of the RSOC is smaller than the hopping integral $t$
and the odd-frequency spin-triplet $s$-wave pair amplitude increases with
the increase of the amplitude of the RSOC\@.
On the other hand, when the amplitude of the RSOC is much larger than the hopping integral,
the ZESABS disappears
when the value of the chemical potential is 
non-zero
as can be seen in the Appendix A
and the odd-frequency spin-triplet $s$-wave pair amplitude disappears as increasing
$\lambda/t$ for the large amplitude of the RSOC\@.
\par

A ZESABS is also produced by a Dresselhaus spin-orbit interaction and
the Zeeman field~\cite{Ikegaya2016}.
The Dresselhaus spin-orbit interaction breaks the spin-rotational symmetry
and the anomalous proximity effect is also expected for the spin-singlet
$d_{xy}$-wave superconductors.
\par

Recently, relation between crossed Andreev reflection (CAR) and odd frequency pair
amplitude is pointed out~\cite{Crepin}.
On the other hand, in ballistic transport regime, CAR
with mixed parity state is calculated~\cite{0953-8984-29-1-015301}.
It is an interesting future problem to calculate CAR in diffusive regime.

In this paper, we focus on the localized 
odd-frequency pairing accompanied by the ZESABS, which is generated by the 
symmetry breaking from the bulk conventional even-frequency pair potential. 
On the other hand, several studies have been
pursuing the realization of a bulk odd-frequency superconductor 
so far~\cite{Berezinskii,Belitz1,Belitz2,Balatsky,Emery,Coleman2,Balatsky2,
   Coleman,Belitz3,Vojta,Fuseya,Shigeta,Hotta,Shigeta2,Solenov,
Fuseya2011,Kusunose,Kusunose2,Fukui}. 
Although it is difficult to express the bulk odd-frequency superconducting 
states consistent with various conditions~\cite{Asano2014,Fominov2015}, 
a recently proposed odd-frequency gap function in the two channel Kondo lattice model 
(TCKL) is a promising system~\cite{Hoshino2015}. 
The calculation of charge transport and Josephson 
effect has already started~\cite{Hoshino2016}. 
It is interesting to study proximity effects in the
DN/TCKL system.

\par

\begin{acknowledgments}
   Y. Asano, S. Kobayashi S. Nakosai are thanked for useful discussions.
   This work was supported
   by Grants-in-Aid from JSPS
   for Scientific Research on Innovative Areas “Topological Materials
   Science” (KAKENHI Grant Numbers  JP15H05851, JP15H05853 and JP15K21717),
   for Scientific Research (B) (KAKENHI Grant Numbers JP15H03686 and JP18H01176),
   and Japan-RFBR JSPS Bilateral Joint Research Projects/Seminars.
\end{acknowledgments}

\bibliography{./TopologicalSC.bib}
\appendix
\begin{widetext}

\section{Derivation of surface Green's function\label{sec:App_der_green}}

In this Appendix, we explain 
a recursive method to calculate the Green's function~\cite{Umerski}.
The Hamiltonian with open boundary condition in the $x$-direction
and periodic boundary condition in the $y$-direction is given by
\begin{align}
   {\cal H}_{n,\mathrm{SC}}
   =&
   \sum_{k_y}
   C_{n,k_y}^\dag
   H_n(k_y)
   C_{n,k_y},
   \\
   H_n(k_y)
   =&
   \begin{pmatrix}
      \hat{u} & \hat{t} &
      \\
      \hat{t}^\dag & \hat{u} &\hat{t}&
      \\
      &\hat{t}^\dag&\ddots&\ddots
      \\
      &&&\ddots&\hat{t}
      \\
       &&&\hat{t}^\dag&\hat{u}
   \end{pmatrix},
   \\
   \hat{u}
   =&
   \begin{pmatrix}
      \varepsilon_{k_y} & \lambda\sin k_y
      \\
      \lambda\sin k_y & \varepsilon_{k_y}
      \\
                      &&-\varepsilon_{k_y} & \lambda\sin k_y
                      \\
                      &&\lambda\sin k_y & -\varepsilon_{k_y}
   \end{pmatrix},
   \\
   {\hat{t}} =&
   \begin{pmatrix}
      -t & \lambda/2 &0&\Delta_{12}(k_y)
      \\
      -\lambda/2&-t&\Delta_{21}(k_y) &0
      \\
      0          &\Delta_{21}(k_y)&t &-\lambda/2
                \\
      \Delta_{12}(k_y)&0&\lambda/2&t
   \end{pmatrix},
\end{align}
where 
   $n$ is the number of sites in the $x$-direction,
$\varepsilon_{k_y}=-2t\cos k_y-\mu$,
$\Delta_{12}$ and 
$\Delta_{21}$ are given by
\begin{align}
   \Delta_{12}
   (k_y)
   =&
   \Delta_0/(2i),
   \\
   \Delta_{21}
   (k_y)
   =&
   \Delta_0/(2i),
\end{align}
for the spin-triplet $p_x$-wave case,
$\Delta_{12}$ and 
$\Delta_{21}$ are given by
\begin{align}
   \Delta_{12}
   (k_y)
   =&
   -i\Delta_{0}\sin k_y/2,
   \\
   \Delta_{21}
   (k_y)
   =&
   i\Delta_{0}\sin k_y/2,
\end{align}
for the spin-singlet $d_{xy}$-wave case
and
\begin{align}
   C_{n,k_y}
   =
   &
   \left(
      c_{1,k_y,\uparrow},\:
      c_{1,k_y,\downarrow},\:
      c_{1,-k_y,\uparrow}^\dag,\:
      c_{1,-k_y,\downarrow}^\dag,
      \ldots,
      c_{n,k_y,\uparrow},\:
      c_{n,k_y,\downarrow},\:
      c_{n,-k_y,\uparrow}^\dag
      c_{n,-k_y,\downarrow}^\dag
   \right)
   ^{\mathrm{T}}.
\end{align}

Let M\"obius transformation be
\begin{align}
   \hat{A}_\bullet\hat{z}
   =&
   \left(
      \hat{a}\hat{z}
      +
      \hat{b}
   \right)
   {\left(
         \hat{c}\hat{z}
         +
         \hat{d}
   \right)}^{-1},
   \label{eq:Mobius}
\end{align}
with
\begin{align}
   \hat{A}
   =&
   \begin{pmatrix}
      \hat{a} & \hat{b}
      \\
      \hat{c} & \hat{d}
   \end{pmatrix},
\end{align}
where $\hat{a}$, $\hat{b}$, $\hat{c}$, $\hat{d}$ and $\hat{z}$ are 
$m\times m$ matrices.
The M\"obius transformation satisfies following relation:
\begin{align}
   \hat{A}_\bullet
   \hat{B}_\bullet
   \hat{z}
   =
   {%
      \left(
         \hat{A}
         \hat{B}
      \right)
   }_\bullet
   \hat{z}.
\end{align}
We also define the elements of the Green's function 
as
\begin{align}
   {%
      \left[
         z-H_n(k_y)
      \right]
   }^{-1}
   =&
   \begin{pmatrix}
      {[g_n(z,k_y)]}_{1,1} & {[g_n(z,k_y)]}_{1,2} & \ldots
      \\
      {[g_n(z,k_y)]}_{2,1} & {[g_n(z,k_y)]}_{2,2} &
      \\
      \vdots &&\ddots
   \end{pmatrix},
\end{align}
with $4\times4$ matrix ${[g_n(z,k_y)]}_{j_x,j_x'}$.
Then the recurrence relation for the surface Green's function for the $n$-site system 
at the leftmost site ${[g_n(z,k_y)]}_{1,1}$
and the $n+1$-site one ${[g_{n+1}(z,k_y)]}_{1,1}$ is
given by
\begin{align}
   {%
      \left[
         g_{n+1}(z,k_y)
      \right]
   }_{1,1}
   =&
      {%
         \left\{
            \left(
               z-\hat{u}
            \right)
            -
            \hat{t}
            {%
               \left[
                  g_n(z,k_y)
               \right]
            }_{1,1}
            \hat{t}^\dag
         \right\}
      }^{-1}
   \nonumber\\
   =&
   X_\bullet
   {%
      \left[
         g_{n}(z,k_y)
      \right]
   }_{1,1},
\end{align}
with
\begin{align}
   X
   =&
   \begin{pmatrix}
      \hat{0} &{(\hat{t}^\dag)}^{-1}
      \\
      -\hat{t} & (z-\hat{u}){(\hat{t}^\dag)}^{-1}
   \end{pmatrix}.
\end{align}
Here, $\hat{0}$ is a $4\times4$ matrix with all the elements equal to zero.
   It is noted ${[g_1(z,k_y)]}_{1,1}$ satisfies
   \begin{align}
      {[g_1(z,k_y)]}_{1,1}
      =&
      X_\bullet
      \hat{0},
   \end{align}
   and ${[g_n(z,k_y)]}_{1,1}$ is written as
   \begin{align}
      {[g_n(z,k_y)]}_{1,1}
      =&
      {%
         \left(
            X^n
         \right)
      }_\bullet \hat{0}.
   \end{align}
Let $U$ be a matrix which diagonalizes $X$ as
\begin{align}
   U^{-1}XU
   =&
   \mathrm{diag}
   \left(
      \lambda_1,
      \lambda_2,
      \ldots,
      \lambda_8
   \right),
\end{align}
with
$|\lambda_1|\leq\cdots\leq|\lambda_4|<|\lambda_5|\leq\cdots\leq|\lambda_8|$.
Then the surface Green's function for a semi-infinite system at the leftmost site 
 $G_{j_x=1}^\infty(z,k_y)$ is given by
\begin{align}
   G_{j_x=1}^\infty(z,k_y)
   =&
   \lim_{n\rightarrow\infty}[g_n(z,k_y)]_{1,1}
   \\
   =&
      \lim_{n\rightarrow\infty}
      {%
         \left[
            U
            {%
               \left(
                  U^{-1}XU
               \right)
            }^n
            U^{-1}
         \right]
      }_\bullet
      \hat{0}
   \nonumber\\
   =&
   U_\bullet \hat{0}
   \nonumber\\
   =&
   U_{12}{(U_{22})}^{-1},
   \label{eq:surface_Green}
\end{align}
where 
$U_{ij}$ is a $4\times4$ matrix defined as
\begin{align}
   U
   =&
   \begin{pmatrix}
      U_{11} & U_{12}
      \\
      U_{21} & U_{22}
   \end{pmatrix},
\end{align}
and we use Eq.~(\ref{eq:Mobius}).

\section{Winding number for spin-singlet $d_{xy}$-wave superconductor with RSOC\label{sec:App_topo_dxy}}
From Eq.~(\ref{eq:dispersion_dxy}), point nodes are on $k_{x(y)}=0$ or $k_{x(y)}=\pi$, 
since $\Delta_0\sin k_x\sin k_y=0$ is satisfied at these points.
We consider only $k_x=0$ or $k_x=\pi$ case because Eq.~(\ref{eq:dispersion_dxy}) 
is symmetric with respect to replacing $k_x$ by $k_y$.
Point nodes also satisfy $\xi_k\pm|\lambda(\sin k_x+\sin k_y)|=0$.
For $k_x=0$, point nodes are located at
\begin{align}
   k_y 
   =&
   k_{0,\pm}
   =
   \arccos
   \left\{
      \frac{1}{4t^2+\lambda^2}
      \left[
         -
         2
         t(2t+\mu)
         \pm
         \lambda
         \sqrt{%
            \lambda^2
            -
            4t\mu
            -
            \mu^2
         }%
      \right]
   \right\},
\end{align}
with $\lambda^2 - 4t\mu - \mu^2\geq0$
and for $k_x=\pi$, point nodes are located at
\begin{align}
   k_y 
   =&
   k_{\pi,\pm}
   =
   \arccos
   \left\{
      \frac{1}{4t^2+\lambda^2}
      \left[
         -
         2
         t(-2t+\mu)
         \pm
         \lambda
         \sqrt{%
            \lambda^2
            +
            4t\mu
            -
            \mu^2
         }%
      \right]
   \right\},
\end{align}
with $\lambda^2 + 4t\mu - \mu^2\geq0$.
Here we suppose $\lambda\geq0$.
Then the number of point nodes is
\begin{align}
   \begin{cases}
      8 & -2t-\sqrt{4t^2+\lambda^2}<\mu<+2t-\sqrt{4t^2+\lambda^2},
      \\
      16 & +2t-\sqrt{4t^2+\lambda^2}<\mu<-2t+\sqrt{4t^2+\lambda^2},
      \\
      8 & -2t+\sqrt{4t^2+\lambda^2}<\mu<+2t+\sqrt{4t^2+\lambda^2},
   \end{cases}
\end{align}
with the 
non-zero
value of the RSOC\@.

As shown in Fig.~\ref{fig:phase_diag}, there are nine regimes from A to I.
\begin{figure}[htbp]
   \centering
   \includegraphics[width = 8.6cm]{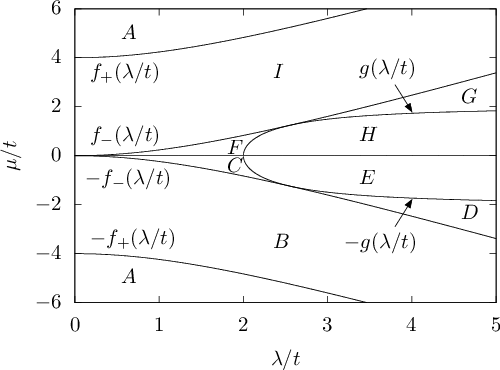}
   \caption{%
      Zero-energy flatband is characterized by the region A to I.
      $f_\pm(\lambda/t)=\pm 2+\sqrt{4^2+{(\lambda/t)}^2}$, and
      $g(\lambda/t)=\frac{2}{\lambda/t}\sqrt{{(\lambda/t)}^2-4^2}$.
   }%
   \label{fig:phase_diag}
\end{figure}
The region A in Fig.~\ref{fig:phase_diag}, the winding number is zero for any $k_y$.

The winding number for $-2t-\sqrt{4t^2+\lambda^2}<\mu<+2t-\sqrt{4t^2+\lambda^2}$
[region B in Fig.~\ref{fig:phase_diag}]
is
\begin{align}
   W(k_y)
   =&
   \begin{cases}
      0 & -\pi<k_y<-k_{0,-},
      \\
      -1 & -k_{0,-}<k_y<-k_{0,+},
      \\
      -2 & -k_{0,+}<k_y<0,
      \\
      2 & 0<k_y<k_{0,+},
      \\
      1 & k_{0,+}<k_y<k_{0,-},
      \\
      0 & k_{0,-}<k_y<\pi.
   \end{cases}
\end{align}
The winding number 
for $+2t-\sqrt{4t^2+\lambda^2}<\mu<0$ and $\lambda<\sqrt{2(1+\sqrt{5})}t$ 
[region C in Fig.~\ref{fig:phase_diag}]
is
\begin{align}
   W(k_y)
   =&
   \begin{cases}
      0 & -\pi<k_y<-k_{0,-},
      \\
      -1 & -k_{0,-}<k_y<-k_{0,+},
      \\
      -2 & -k_{0,+}<k_y<-k_{\pi,-},
      \\
      -1 & -k_{\pi,-}<k_y<-k_{\pi,+},
      \\
      -2 & -k_{\pi,+}<k_y<0,
      \\
      2 & 0<k_y<k_{\pi,+},
      \\
      1 & k_{\pi,+}<k_y<k_{\pi,-},
      \\
      2 & k_{\pi,-}<k_y<k_{0,+},
      \\
      1 & k_{0,+}<k_y<k_{0,-},
      \\
      0 & k_{0,-}<k_y<\pi.
   \end{cases}
\end{align}
The winding number 
for $+2t-\sqrt{4t^2+\lambda^2}<\mu<-\frac{2t}{\lambda}\sqrt{\lambda^2-4t^2}$ 
and $\lambda>\sqrt{2(1+\sqrt{5})}t$ 
[region D in Fig.~\ref{fig:phase_diag}]
is
\begin{align}
   W(k_y)
   =&
   \begin{cases}
      0 & -\pi<k_y<-k_{0,-},
      \\
      -1 & -k_{0,-}<k_y<-k_{\pi,-},
      \\
      0 & -k_{\pi,-}<k_y<-k_{\pi,+},
      \\
      -1 & -k_{\pi,+}<k_y<-k_{0,+},
      \\
      -2 & -k_{0,+}<k_y<0,
      \\
      2 & 0<k_y<k_{0,+},
      \\
      1 & k_{0,+}<k_y<k_{\pi,+},
      \\
      0 & k_{\pi,+}<k_y<k_{\pi,-},
      \\
      1 & k_{\pi,-}<k_y<k_{0,-},
      \\
      0 & k_{0,-}<k_y<\pi.
   \end{cases}
\end{align}
The winding number 
for $-\frac{2t}{\lambda}\sqrt{\lambda^2-4t^2}<\mu<0$ 
[region E in Fig.~\ref{fig:phase_diag}]
is
\begin{align}
   W(k_y)
   =&
   \begin{cases}
      0 & -\pi<k_y<-k_{0,-},
      \\
      -1 & -k_{0,-}<k_y<-k_{\pi,-},
      \\
      0 & -k_{\pi,-}<k_y<-k_{0,+},
      \\
      -1 & -k_{0,+}<k_y<-k_{\pi,+},
      \\
      -2 & -k_{\pi,+}<k_y<0,
      \\
      2 & 0<k_y<k_{\pi,+},
      \\
      1 & k_{\pi,+}<k_y<k_{0,+},
      \\
      0 & k_{0,+}<k_y<k_{\pi,-},
      \\
      1 & k_{\pi,-}<k_y<k_{0,-},
      \\
      0 & k_{0,-}<k_y<\pi.
   \end{cases}
\end{align}
The winding number for $0<\mu<-2t+\sqrt{4t^2+\lambda^2}$ and $\lambda<\sqrt{2(1+\sqrt{5})}t$
[region F in Fig.~\ref{fig:phase_diag}]
is
\begin{align}
   W(k_y)
   =&
   \begin{cases}
      -2 & -\pi<k_y<-k_{0,-},
      \\
      -1 & -k_{0,-}<k_y<-k_{0,+},
      \\
      -2 & -k_{0,+}<k_y<-k_{\pi,-},
      \\
      -1 & -k_{\pi,-}<k_y<-k_{\pi,+},
      \\
      0 & -k_{\pi,+}<k_y<k_{\pi,+},
      \\
      1 & k_{\pi,+}<k_y<k_{\pi,-},
      \\
      2 & k_{\pi,-}<k_y<k_{0,+},
      \\
      1 & k_{0,+}<k_y<k_{0,-},
      \\
      2 & k_{0,-}<k_y<\pi.
   \end{cases}
\end{align}
The winding number for $0<\mu<-2t+\sqrt{4t^2+\lambda^2}$ and $\lambda>\sqrt{2(1+\sqrt{5})}t$
[region G in Fig.~\ref{fig:phase_diag}]
is
\begin{align}
   W(k_y)
   =&
   \begin{cases}
      -2 & -\pi<k_y<-k_{\pi,-},
      \\
      -1 & -k_{\pi,-}<k_y<-k_{0,-},
      \\
      0 & -k_{0,-}<k_y<-k_{0,+},
      \\
      -1 & -k_{0,+}<k_y<-k_{\pi,+},
      \\
      0 & -k_{\pi,+}<k_y<k_{\pi,+},
      \\
      1 & k_{\pi,+}<k_y<k_{0,+},
      \\
      0 & k_{0,+}<k_y<k_{0,-},
      \\
      1 & k_{0,-}<k_y<k_{\pi,-},
      \\
      2 & k_{\pi,-}<k_y<\pi.
   \end{cases}
\end{align}
The winding number for $0<\mu<\frac{2t}{\lambda}\sqrt{\lambda^2-4t^2}$
[region H in Fig.~\ref{fig:phase_diag}]
is
\begin{align}
   W(k_y)
   =&
   \begin{cases}
      -2 & -\pi<k_y<-k_{0,-},
      \\
      -1 & -k_{0,-}<k_y<-k_{\pi,-},
      \\
      0 & -k_{\pi,-}<k_y<-k_{0,+},
      \\
      -1 & -k_{0,+}<k_y<-k_{\pi,+},
      \\
      0 & -k_{\pi,+}<k_y<k_{\pi,+},
      \\
      1 & k_{\pi,+}<k_y<k_{0,+},
      \\
      0 & k_{0,+}<k_y<k_{\pi,-},
      \\
      1 & k_{\pi,-}<k_y<k_{0,-},
      \\
      2 & k_{0,-}<k_y<\pi.
   \end{cases}
\end{align}
The winding number for $-2t+\sqrt{4t^2+\lambda^2}<\mu<+2t+\sqrt{4t^2+\lambda^2}$
[region I in Fig.~\ref{fig:phase_diag}]
is
\begin{align}
   W(k_y)
   =&
   \begin{cases}
      -2 & -\pi<k_y<-k_{\pi,-},
      \\
      -1 & -k_{\pi,-}<k_y<-k_{\pi,+},
      \\
      0 & -k_{\pi,+}<k_y<k_{\pi,+},
      \\
      1 & k_{\pi,+}<k_y<k_{\pi,-},
      \\
      2 & k_{\pi,-}<k_y<\pi.
   \end{cases}
\end{align}
\section{%
   Odd-frequency spin-triplet $s$-wave pair amplitude for semi-infinite system%
   \label{sec:App_odd_pair}
}
\begin{figure*}[htbp]
   \centering
   \includegraphics[width = 17cm]{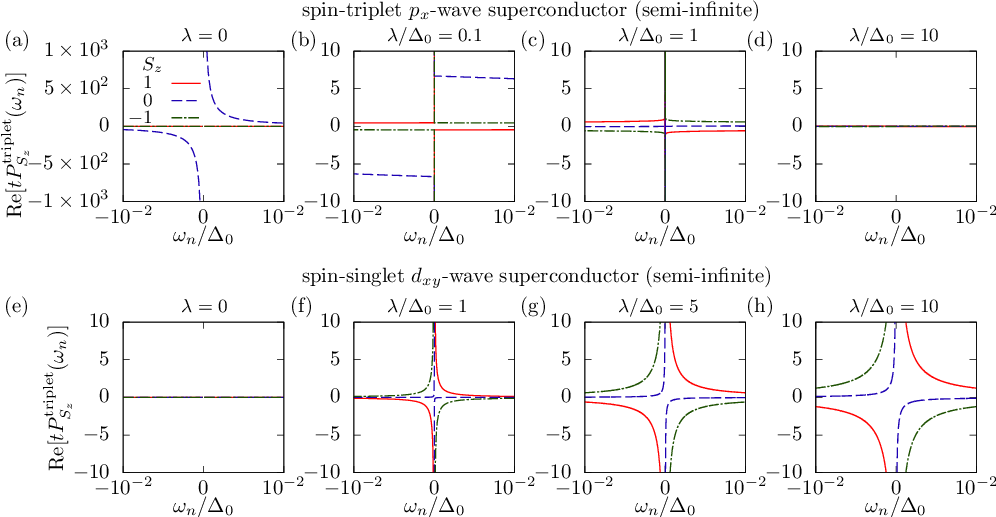}
   \caption{%
      The real part of the odd-frequency spin-triplet $s$-wave pair amplitude 
      $P_{S_z}^\mathrm{triplet}(\omega_n)$ 
      is plotted as a function of $\omega_n$ for $S_z=-1$, $0$, $1$.
      The imaginary part of $P_{S_z}^\mathrm{triplet}(\omega_n)$ is zero within numerical accuracy.
      The odd-frequency pair amplitude of the spin-triplet $p_x$-wave superconductor is
      shown in 
      (a) $\lambda/\Delta_0=0$, (b) $0.1$, (c) $1$ and (d) $10$.
      The odd-frequency pair amplitude of the spin-singlet $d_{xy}$-wave superconductor is
      shown in 
      (e) $\lambda/\Delta_0=0$, (f) $1$, (g) $5$ and (h) $10$. In all cases, we set $\Delta_0/t=0.01$.
   }%
   \label{fig:pair_ky}
\end{figure*}
Frequency dependence of the odd-frequency spin-triplet $s$-wave pair amplitude 
at the surface of the semi-infinite SC is plotted for $S_z=1$, $0$ and $-1$ in Fig.~\ref{fig:pair_ky}.

In Figs.~\ref{fig:pair_ky} (a) to (d), the odd-frequency spin-triplet $s$-wave pair amplitude
for the spin-triplet $p_x$-wave superconductor are shown.
Without the RSOC, $P^\mathrm{triplet}_{S_z}(\omega_n)$ with $S_z=\pm1$ is zero
due to the spin-rotational symmetry but $P^\mathrm{triplet}_{S_z}(\omega_n)$ with $S_z=0$
has large value due to 
translational symmetry braking
and it diverges 
near
$\omega_n=0$ [Fig.~\ref{fig:pair_ky}~(a)].
At $\lambda/\Delta_0=0.1$ shown in Fig.~\ref{fig:pair_ky}~(b), $P^\mathrm{triplet}_{S_z}(\omega_n)$ with $S_z=0$
greatly suppressed and $P^\mathrm{triplet}_{S_z}(\omega_n)$ with $S_z=\pm1$
has 
non-zero
value due to the spin-rotational symmetry breaking.
$P^\mathrm{triplet}_{S_z}(\omega_n)$ diverges very close to $k_y=0$ since
there is ZESABS at $k_y=0$.
For $\lambda/\Delta_0\gtrsim1$, $P^\mathrm{triplet}_{S_z}(\omega_n)$
with $S_z=0$ almost vanishes [Figs.~\ref{fig:pair_ky}~(c) and (d)]
and all the components of $P^\mathrm{triplet}_{S_z}(\omega_n)$ are
almost zero for $\lambda/\Delta_0\gtrsim10$ shown in Fig.~\ref{fig:pair_ky}~(d).
For $\lambda/\Delta_0=10$, we cannot see divergent behavior for
$|\omega_n/\Delta_0|>5\times10^{-6}$.

In Figs.~\ref{fig:pair_ky} (e) to (h), the odd-frequency spin-triplet $s$-wave 
pair amplitude
for the spin-singlet $d_{xy}$-wave superconductor are shown.
$P^\mathrm{triplet}_{S_z}(\omega_n)$ is zero for $\lambda/\Delta_0=0$ due to
the spin-rotational symmetry [Fig.~\ref{fig:pair_ky}~(e)].
For $\lambda/\Delta_0>0$, absolute value of $P^\mathrm{triplet}_{S_z}(\omega_n)$
increases as the increase of $\lambda/\Delta_0$ [Figs.~\ref{fig:pair_ky}~(f) to (h)].
The absolute value of $P^\mathrm{triplet}_{S_z}(\omega_n)$ with $S_z=\pm1$ is
larger than that with $S_z=0$ since the even-frequency spin-triplet with $S_z=0$
pair amplitude is absent in the bulk [see also the Appendix~\ref{sec:App_Green_dxy}].
In these graphs, the absolute value of $P^\mathrm{triplet}_{S_z}(\omega_n)$
increases as the increase of $\lambda$.

\section{%
   Odd-frequency spin-singlet $p_y$-wave pair amplitude
   for spin-singlet $d_{xy}$-wave superconductor%
\label{sec:App_dxy_py}
}
In Fig.~\ref{fig:omega_dep_dxy_py}, the odd-frequency spin-singlet $p_y$-wave pair amplitude
at the surface
is shown for the semi-infinite spin-singlet $d_{xy}$-wave superconductor.
The definition of the odd-frequency spin-singlet $p_y$-wave pair amplitude is given by
\begin{align}
   P^\mathrm{singlet}_{p_y}(\omega_n)
   =&
   \frac{1}{2N_{k_y}}
   \sum_{k_y}
   \left[
      \hat{F}_{j_x=1,\uparrow,\downarrow}(i\omega_n,k_y)
      -
      \hat{F}_{j_x=1,\downarrow,\uparrow}(i\omega_n,k_y)
   \right]
   \sin k_y,
\end{align}
where $\hat{F}_{j_x,\sigma,\sigma'}(i\omega_n,k_y)$ is the anomalous Green's function
given by Eq.~(\ref{eq:anomalous_G}).
The amplitude does not almost depend on the value of $\lambda$.

\begin{figure}[htbp]
   \centering
   \includegraphics[width = 16cm]{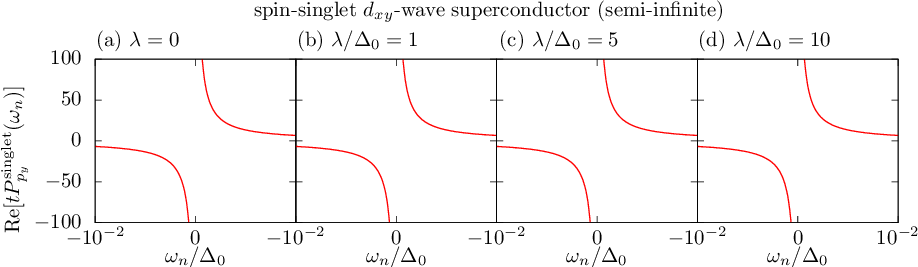}
   \caption{%
         The real part of
      the odd-frequency spin-singlet $p_y$-wave pair amplitude
      is plotted as a function of $\omega_n/\Delta_0$ for 
      (a) $\lambda/\Delta_0=0$,
      (b) $1$,
      (c) $5$, and
      (d) $10$ with $\Delta_0/t=0.01$.
      The imaginary part of the odd-frequency 
      spin-singlet $p_y$-wave pair amplitude is zero within numerical accuracy.
   }%
   \label{fig:omega_dep_dxy_py}
\end{figure}

\section{%
   Anomalous Green's function for $d_{xy}$-wave superconductor for bulk system%
   \label{sec:App_Green_dxy}
}
The bulk Hamiltonian with $d_{xy}$-wave superconductor is given by
\begin{align}
   H_\mathrm{bulk}
   =&
   \sum_\mathbf{k}
   \bar{C}_\mathbf{k}^\dag
   H(\mathbf{k})
   \bar{C}_\mathbf{k},
   \\
   H(\mathbf{k})
   =&
   \begin{pmatrix}
      E(\mathbf{k}) & \Delta(\mathbf{k})  
      \\
      \Delta^\dag(\mathbf{k}) & E^\mathrm{T}(-\mathbf{k})
   \end{pmatrix},
   \\
   E(\mathbf{k})
   =&
   E_0\sigma_0
   +
   E_1\sigma_1
   +
   E_2\sigma_2,
   \\
   E_0
   =&
   \left[
      -2t
      \left(
         \cos k_x + \cos k_y
      \right)
      -\mu
   \right]
   \sigma_0,
   \\
   E_1
   =&
   +
   \lambda\sin k_y\sigma_1,
   \\
   E_2
   =&
   \lambda\sin k_x\sigma_2,
   \\
   \Delta(\mathbf{k})
   =&
   \Delta_0\sin k_x\sin k_y i\sigma_2
   \\
   =&
   \tilde{\Delta}\sigma_2,
   \\
   \bar{C}_\mathrm{k}
   =&
   \left(
      c_{\uparrow,\mathbf{k}},
      c_{\downarrow,\mathbf{k}},
      c_{\uparrow,-\mathbf{k}}^\dag,
      c_{\downarrow,-\mathbf{k}}^\dag
   \right).
\end{align}
The anomalous Green's function for bulk system with the spin-singlet $d_{xy}$-wave superconductor is given by
\begin{align}
   G_{12}(z,\mathbf{k})
   =&
   \frac{1}{\alpha_0^2-\alpha_2^2-\alpha_3^2}
   \left(
      \alpha_0\sigma_0
      -
      \alpha_2\sigma_2
      -
      \alpha_3\sigma_3
   \right),
   \label{eq:app_G_12}
   \\
   \alpha_0
   =&
   \frac{2E_0E_2}{\tilde{\Delta}},
   \\
   \alpha_2
   =&
   -\tilde{\Delta}^*
   +
   \frac{z^2}{\tilde{\Delta}}
   -
   \frac{1}{\tilde{\Delta}}
   \left(
      E_0^2+E_1^2+E_2^2
   \right),
   \\
   \alpha_3
   =&
   -\frac{2iE_0E_1}{\tilde{\Delta}},
\end{align}
with
\begin{align}
   G(z,\mathbf{k})
   =&
   {%
      \left[
         z-H(\mathbf{k})
      \right]
   }^{-1},
   \\
   G(z,\mathbf{k})
   =&
   \begin{pmatrix}
      G_{11}(z,\mathbf{k})
      &
      G_{12}(z,\mathbf{k})
      \\
      G_{21}(z,\mathbf{k})
      &
      G_{22}(z,\mathbf{k})
   \end{pmatrix},
\end{align}
where $G_{ij}(z,\mathbf{k})$ ($i,j=1,2$) is a $2\times2$ matrix.
From Eq.~(\ref{eq:app_G_12}), we can see that the spin-triplet component with $S_z=0$ 
is absent since $S_z=0$ component is proportional to $\sigma_1$.

\section{Matrix elements of Green's function for SC/DN junction\label{sec:App_H_DN_SC}}
In this Appendix, we explain the derivation of the Green's function in the
DN in DN/superconductor junctions
where we use super unit cell in the $y$-direction 
($-\pi/L_y<k_y\leq \pi/L_y$) with periodic boundary condition.
The Hamiltonian in the DN is
\begin{align}
   {\cal H}_\mathrm{DN}^{l_s}
   =&
   \sum_{k_y}
   \tilde{C}^\dag(k_y)
   \begin{pmatrix}
      \tilde{u}^{(-49),l_s} & \tilde{t}
      \\
      \tilde{t}^\dag & \tilde{u}^{(-48),l_s} & \tilde{t}
      \\
                     &\tilde{t}^\dag & \tilde{u}^{(-47),l_s} & \tilde{t}
                      \\
                      & &&\ddots&
                      \\
                      &&& && \tilde{t}
                      \\
                      &&&&\tilde{t}^\dag & \tilde{u}^{(0),l_s}
   \end{pmatrix}
   \tilde{C}(k_y),
\end{align}
with $4L_y\times 4L_y$ matrices
\begin{align}
   \tilde{u}^{(j_x),l_s}
   =&
   \begin{pmatrix}
      A_{1}^{(j_x,1),l_s} & A_{2} & \hat{0} &\cdots & \hat{0} & A_{2}^\dag
      \\
      A_{2}^\dag & A_{1}^{(j_x,2),l_s} & A_{2} & \hat{0} && \hat{0}
      \\
      \hat{0}          &A_{2}^\dag & A_{1}^{(j_x,3),l_s} & A_{2}&\hat{0}&\vdots
      \\
      \vdots &&&\ddots&&\hat{0}
      \\
      \hat{0}&&\hat{0}&A_{2}^\dag & A_{1}^{(j_x,L_y-1),l_s}& A_{2}
      \\
      A_{2}&\hat{0}&\cdots &\hat{0}&A_{2}^\dag & A_{1}^{(j_x,L_y),l_s}
   \end{pmatrix},
   \\
   \tilde{t}
   =&
   \begin{pmatrix}
      B_{1} & B_{12} & \hat{0} & \cdots & \hat{0} &B_{21}
      \\
      B_{21} & B_{1} & B_{12} & \hat{0} & & \hat{0}
      \\
      \hat{0}   &B_{21} & B_{1} & B_{12}& \hat{0} &\vdots
      \\
      \vdots &&&\ddots&& \hat{0}
      \\
      \hat{0}      &\cdots & \hat{0} &B_{21} & B_{1}&B_{12}
      \\
      B_{12}& \hat{0} &\cdots & \hat{0} &B_{21} & B_{1}
   \end{pmatrix},
\end{align}
and
\begin{align}
   \tilde{C}(k_y)
   =&
   {%
      \left(
         \tilde{\boldsymbol{c}}_{(-49,1),k_y},
         \tilde{\boldsymbol{c}}_{(-49,2),k_y},
         \cdots,
         \tilde{\boldsymbol{c}}_{(-49,L_y),k_y},
         \tilde{\boldsymbol{c}}_{(-48,1),k_y},
         \tilde{\boldsymbol{c}}_{(-48,2),k_y},
         \cdots,
         \tilde{\boldsymbol{c}}_{(0,L_y),k_y},
      \right)
   }^\mathrm{T},
   \\
   \tilde{\boldsymbol{c}}_{\mathbf{j},k_y}
   =&
   \left(
      c_{\mathbf{j},k_y,\uparrow},
      c_{\mathbf{j},k_y,\downarrow},
      c_{\mathbf{j},-k_y,\uparrow}^\dag,
      c_{\mathbf{j},-k_y,\downarrow}^\dag
   \right).
\end{align}
Here, 
$A_{1}^{(j_x,y_y),l_s}$ and $A_{2}$ are $4\times 4$ matrices given by
\begin{align}
   A_{1}^{(j_x,j_y),l_s}
   =&
   \begin{pmatrix}
      -\mu+V_{(j_x,j_y)}^{l_s}
      \\
      &-\mu+V_{(j_x,j_y)}^{l_s}
      \\
      &&\mu-V_{(j_x,j_y)}^{l_s}
      \\
      &&&\mu-V_{(j_x,j_y)}^{l_s}
   \end{pmatrix},
   \\
   A_2
   =&
   \begin{pmatrix}
      -te^{ik_y} & -i\lambda e^{ik_y}/2
      \\
      -i\lambda e^{ik_y}/2& -te^{ik_y}
      \\
                                 &&te^{ik_y}&-i\lambda e^{ik_y}/2
      \\
      &&-i\lambda e^{ik_y}/2& te^{ik_y}
   \end{pmatrix},
\end{align}
where $-t<V_\mathbf{j}^{l_s}<t$ is a random potential at $\mathbf{j}=(j_x,j_y)$-th site.
$B_1$, $B_{12}$ and $B_{21}$ are $4\times 4$ matrices and they are given by
\begin{align}
   B_1 
   =&
   \begin{pmatrix}
      -t &\lambda/2&&-i\Delta_0/2
      \\
      -\lambda/2&-t&-i\Delta_0/2
         \\
      &-i\Delta_0/2&t&-\lambda/2
      \\
      -i\Delta_0/2&&\lambda/2&t
   \end{pmatrix},
   \\
   B_{12}
   =&
   B_{21}
   =
   \hat{0},
\end{align}
for the spin-triplet $p_x$-wave superconductor.
For the spin-singlet $d_{xy}$-wave superconductor, $B_1$,$B_{12}$ and $B_{21}$ are given by
\begin{align}
   B_1 
   =&
   \begin{pmatrix}
      -t &\lambda/2&&
      \\
      -\lambda/2&-t&
         \\
      &&t&-\lambda/2
      \\
      &&\lambda/2&t
   \end{pmatrix},
   \\
   B_{12}
   =&
   -B_{21}^*
   =
   \begin{pmatrix}
      &&&-\Delta_0 e^{ik_y}/4
      \\
      &&\Delta_0 e^{ik_y}/4
      \\
      &\Delta_0 e^{ik_y}/4
      \\
      -\Delta_0 e^{ik_y}/4
   \end{pmatrix}.
\end{align}

The Green's function in the DN is then obtained as
\begin{align}
   \tilde{g}_{j_x}^{\mathrm{DN},l_s}(z,j_y,j'_y,k_y)
   =&
   {%
      \left\{
         {%
            \left[
               \tilde{g}^{(\mathrm{L}),l_s}_{j_x}(z,j_y,j'_y,k_y)
            \right]
         }^{-1}
         -
         \tilde{t}\tilde{g}^{(\mathrm{R}),l_s}_{j_x+1}(z,j_y,j'_y,k_y)\tilde{t}^\dag
      \right\}
   }^{-1},
   \label{eq:tilde_g}
\end{align}
where $\tilde{g}^{(\mathrm{L}),l_s}_{j_x}(z,j_y,j'_y,k_y)$ and 
$\tilde{g}^{(\mathrm{R}),l_s}_{j_x+1}(z,j_y,j'_y,k_y)$ are given by
\begin{align}
   \tilde{g}^{(\mathrm{L}),l_s}_{j_x}(z,j_y,j'_y,k_y)
   =&
   {%
      \left(
         X_{l_x=j_x}^{l_s}
         \ldots
         X_{l_x=-48}^{l_s}
         X_{l_x=-49}^{l_s}
      \right)
   }_\bullet \tilde{0},
   \\
   \tilde{g}^{(\mathrm{R}),l_s}_{j_x+1}(z,j_y,j'_y,k_y)
   =&
   {\left(
         Y_{l_x=j_x+1}^{l_s}
         Y_{l_x=j_x+2}^{l_s}
         \ldots
         Y_{l_x=0}^{l_s}
   \right)}_\bullet \tilde{G}^\infty(z,j_y,j'_y,k_y),
\end{align}
with $4L_y\times 4L_y$ matrix $\tilde{0}$ where all the elements of $\tilde{0}$ are zero.
Here $X_{l_x}^{l_s}$ and $Y_{l_x}^{l_s}$ are given by
\begin{align}
   X_{l_x}^{l_s}
   =&
   \begin{pmatrix}
      \tilde{0} & \tilde{t}^{-1}
      \\
      -\tilde{t}^\dag &[z-\tilde{u}^{(l_x),l_s}]\tilde{t}^{-1}
   \end{pmatrix},
   \\
   Y_{l_x}^{l_s}
   =&
   \begin{pmatrix}
      \tilde{0} & {(\tilde{t}^\dag)}^{-1}
      \\
      -\tilde{t} &[z-\tilde{u}^{(l_x),l_s}]{(\tilde{t}^\dag)}^{-1}
   \end{pmatrix},
\end{align}
and $\tilde{G}^\infty(z,j_y,j'_y,k_y)$ is given by
\begin{align}
   \tilde{G}^\infty(z,j_y,j'_y,k_y)
   =&
   \sum_{m=0,1,\ldots,N_y-1}
   e^{i\frac{2\pi m}{N_y} (j_y-j'_y)}
   G_{j_x=1}^\infty
   \left(
      z,k_y+\frac{2\pi m}{N_y}
   \right),
\end{align}
where $k_y$ is $k_y=-\frac{\pi(N_y-1)}{N_y L_y},-\frac{\pi(N_y-2)}{N_y L_y},\ldots,\frac{\pi}{L_y}$.
$G_{j_x=1}^\infty(z,k'_y)$ with $k'_y=k_y+2\pi m/N_y$ 
is given by Eq.~(\ref{eq:surface_Green}).
\section{Size dependence of LDOS in DN for 
spin-singlet $d_{xy}$-wave superconductor junction\label{sec:App_size}}
In this Appendix, we check the size and the number of sample dependence 
of LDOS at the center of the DN ($j_x=-L_x/2$ where $L_x$ is a length of the DN in the $x$-direction) 
for the spin-singlet $d_{xy}$-wave superconductor junction with $\lambda/\Delta_0=50$.

\begin{figure}[htbp]
   \centering
   \includegraphics[width = 8.5cm]{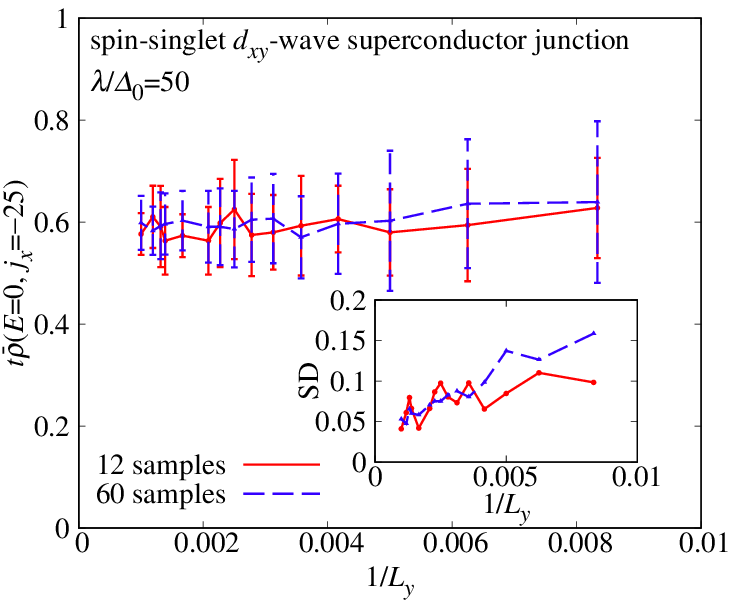}
   \caption{%
      Zero energy LDOS at the center in the DN $\bar{\rho}(E=0,j_x=-25)$ 
      is plotted as a function of the inverse of the size in the $y$-direction $1/L_y$.
      Standard deviation (SD) estimated by 12 and 60 samples is also plotted.
      The length of the DN in the $x$-direction is $L_x=50$.
      SD (error bar in main panel) is plotted in the inset as a function of $1/L_y$.
   }%
   \label{fig:ZEP_size_dxy}
\end{figure}
In Fig.~\ref{fig:ZEP_size_dxy}, we plot LDOS at zero-energy averaged over 12 and 60 samples 
as a function of $1/L_y$.
As we can see in this figure, the extrapolated value of the ZEP for $L_y\rightarrow\infty$ is
much larger than the background value ($\sim0.2$).
In the inset of Fig.~\ref{fig:ZEP_size_dxy}, we plot the corresponding standard deviation.
Although the magnitude of the standard deviation is not small, it decreases with the increase of $L_y$.

\begin{figure}[htbp]
   \centering
   \includegraphics[width = 8.5cm]{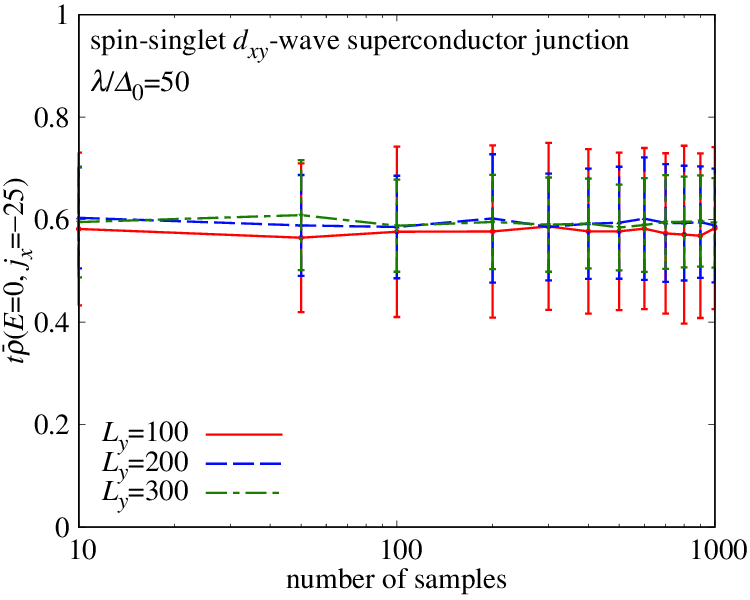}
   \caption{%
      Zero energy LDOS at the center in the DN $\bar{\rho}(E=0,j_x=-25)$ is plotted as a function of 
      the number of samples for $L_y=100$, $200$ and $300$.
      Standard deviation is also plotted.
      The length of the DN is $L_x=50$.
   }%
   \label{fig:ZEP_sample_dxy}
\end{figure}
In Fig.~\ref{fig:ZEP_sample_dxy}, we plot the value of ZEP of LDOS as a function of the number of
samples.
The height of ZEP is almost independent of the number of samples for all values of $L_y$.
The value of standard deviation becomes smaller with the increase of $L_y$.

\begin{figure}[htbp]
   \centering
   \includegraphics[width = 8.5cm]{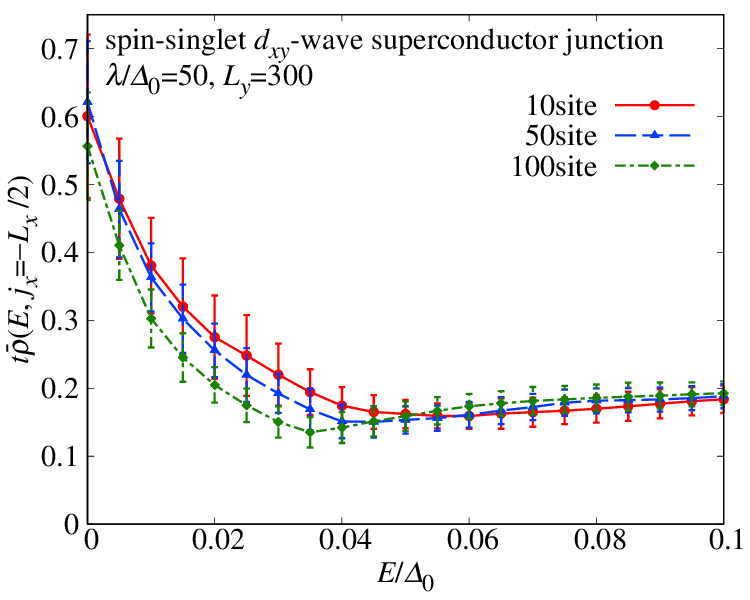}
   \caption{%
      LDOS at the center
      in the DN $\bar{\rho}(E,j_x=-L_x/2)$ for several length of the DN 
      ($L_x=10,50$ and 100)
      is plotted as a function of energy with 60 samples and $L_y=300$.
   }%
   \label{fig:LDOS_size_dxy}
\end{figure}
We also check $L_x$ dependence of LDOS at the center of the DN where $L_x$ is a length 
of the DN in the $x$-direction.
In Fig.~\ref{fig:LDOS_size_dxy}, we change $L_x$ from 10 to 100 and 
plot LDOS at $x=-L_x/2$.
Although, it is difficult to estimate the peak width of LDOS due to the large value of 
standard deviation, it decreases with the increase of $L_x$.

\section{%
   Spatial dependence of odd-frequency pair amplitude
   for the DN/SC junction%
   \label{sec:App_spatial_dep_odd}
}
In this Appendix, we discuss the spatial dependence of the odd-frequency
spin-triplet $s$-wave pair amplitude
for the DN/$d_{xy}$-wave superconductor junction.
The odd-frequency spin-triplet $s$-wave pair amplitudes at each site ($j_x$)
given by
\begin{align}
   P^\mathrm{triplet}_{j_x,S_z=1}(\omega_n)
   =&
   P_{j_x,\uparrow,\uparrow}(\omega_n),
   \label{eq:def_p_s1}
   \\
   P^\mathrm{triplet}_{j_x,S_z=1}(\omega_n)
   =&
   \frac{1}{\sqrt{2}}
   \left[
      P_{j_x,\uparrow,\downarrow}(\omega_n)
      +
      P_{j_x,\downarrow,\uparrow}(\omega_n)
   \right],
   \label{eq:def_p_s0}
   \\
   P^\mathrm{triplet}_{j_x,S_z=-1}(\omega_n)
   =&
   P_{j_x,\downarrow,\downarrow}(\omega_n),
   \label{eq:def_p_s_1}
\end{align}
are shown in Fig.~\ref{fig:DN_SC_onsite}
where $P_{j_x,\sigma,\sigma'}(\omega_n)$ is given by Eq.~(\ref{eq:def_p}).
The odd-frequency spin-singlet $p_y$-wave (onsite) one 
given by
\begin{align}
   P^\mathrm{singlet}_{j_x,p_y}(\omega_n)
   =
   \frac{1}{2N_{k_y}L_y N_\mathrm{sample}}
   \sum_{k_y,j_y,l_s}
   &
   \frac{1}{2\sqrt{2}}
   \left\{
      \left[
         \hat{f}^{l_s}_{j_x,\uparrow,\downarrow}
         (i\omega_n,j_y,j_y+1,k_y)
         -
         \hat{f}^{l_s}_{j_x,\downarrow,\uparrow}
         (i\omega_n,j_y,j_y+1,k_y)
      \right]
   \right.
   \nonumber\\
   &
   \left.
      -
      \left[
         \hat{f}^{l_s}_{j_x,\uparrow,\downarrow}
         (-i\omega_n,j_y,j_y+1,k_y)
         -
         \hat{f}^{l_s}_{j_x,\downarrow,\uparrow}
         (-i\omega_n,j_y,j_y+1,k_y)
      \right]
   \right\}
   e^{-ik_y},
\end{align}
is shown in Fig.~\ref{fig:DN_SC_singlet}.
Here
$\hat{f}^{l_s}_{j_x,\sigma,\sigma'}(z,j_y,j'_y,k_y)$ is the anomalous Green's function of
$\tilde{g}^{\mathrm{DN},ls}_{j_x}(z,j_y,j_y',k_y)$ given by Eq.~(\ref{eq:tilde_g}).

\begin{figure}[h]
   \centering
   \includegraphics[width=18cm]{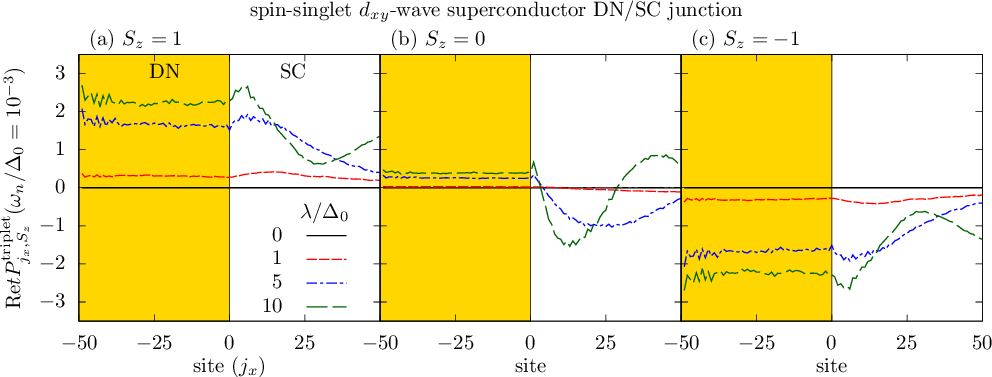}
   \caption{%
      The real part of the odd-frequency spin-triplet $s$-wave pair amplitude 
      with $\omega_n/\Delta_0=10^{-3}$ ($\Delta_0/t=0.01$) is plotted as a function of site ($j_x$)
      for (a) $S_z=1$, (b) $S_z=0$ and (c) $S_z=-1$ with several $\lambda$ ($\lambda/\Delta_0=0$, 1, 5, and 10) 
      for the DN/spin-singlet $d_{xy}$-wave
         superconductor junction.
      The imaginary part of the odd-frequency spin-triplet $s$-wave pair amplitude
      is zero within numerical accuracy.
   }%
   \label{fig:DN_SC_onsite}
\end{figure}
The odd-frequency spin-triplet $s$-wave pair amplitude
$P^\mathrm{triplet}_{j_x,S_z}(\omega_n/\Delta_0=10^{-3})$
is almost constant in the DN and $P_{j_x,S_z=1}(\omega_n)+P_{j_x,S_z=-1}(\omega_n)=0$
holds within numerical accuracy.
The odd-frequency spin-singlet $p_y$-wave pair amplitude 
rapidly decays in the DN since $p$-wave pair amplitudes are fragile against impurity scattering.
$p_x$-wave components are almost zero in the DN and the SC\@.
\begin{figure}[h]
   \centering
   \includegraphics[width=9cm]{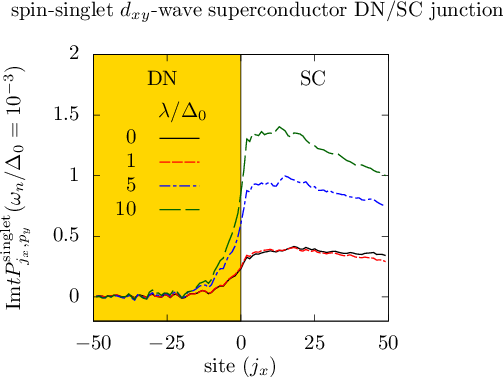}
   \caption{%
      The imaginary part of the odd-frequency spin-singlet $p_y$-wave
      pair amplitude with $\omega_n/\Delta_0=10^{-3}$ ($\Delta_0/t=0.01$) 
      is plotted as a function 
      of site ($j_x$) 
      with several $\lambda$
      ($\lambda/\Delta_0=0$, 1, 5, and 10) 
      for the DN/spin-singlet $d_{xy}$-wave
      superconductor junction.
      The real part of the odd-frequency spin-singlet $p_y$-wave pair amplitude
      is zero within numerical accuracy.
   }%
   \label{fig:DN_SC_singlet}
\end{figure}

\section{%
   Surface roughness for spin-singlet $d_{xy}$-wave superconductor%
   \label{sec:App_tanh}
}
In order to study the effect of roughness, we study a thin diffusive layer 
attached to 
the surface of $d_{xy}$-wave superconductor shown in Fig.~\ref{fig:pic_tanh}. 
The effect of diffusive scattering on the SABS has been studied within quasiclassical approximation~\cite{Nagai96,Golubov1999}. 
Without RSOC, induced odd-frequency pair amplitude is suppressed since 
it has an odd parity~\cite{LuBo2016}. On the other hand, in the presence of 
the RSOC, the odd-frequency spin-triplet $s$-wave pair amplitude is 
induced near the surface%
~\cite{Bakurskiy}. 
Here, we take into account of the spatial depletion of the pair potential near the surface. 
The assumed spatial dependence is shown in Eq.~(\ref{eq:spatila_dep_gap}).
\begin{figure}[htbp]
   \centering
   \includegraphics[width = 7cm]{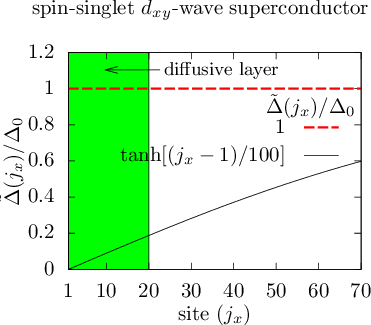}
   \caption{%
      Schematic of the SC with impurity potential ($1\leq j_x\leq N_\mathrm{d}=20$).
      Green colored region indicates the area with impurity potential.
      Pair potential $\tilde{\Delta}(j_x)$ is also plotted as a function of site ($j_x$).
   }%
   \label{fig:pic_tanh}
\end{figure}
The Hamiltonian is given by
\begin{align}
   {\cal H}
   =&
   {\cal H}_t
   +
   \sum_{1\leq j_x\leq N_\mathrm{d},j_y}
   V_{\mathbf{j}}n_\mathbf{j}
   +
   {\cal H}_{\mathrm{SO}}
   -
   \mu\sum_{\mathrm{j}}n_\mathbf{j}
   +
   \tilde{{\cal H}}_{\Delta},
   \\
   \tilde{{\cal H}}_{\Delta}
   =&
   -
   \frac{1}{2}
   \sum_{\mathbf{j}}
   \tilde{\Delta}(j_x)
   \left(
      c_{\mathbf{j},\uparrow}^\dag
      c_{\mathbf{j}+\mathbf{e}_x+\mathbf{e}_y,\downarrow}^\dag
      -
      c_{\mathbf{j},\downarrow}^\dag
      c_{\mathbf{j}+\mathbf{e}_x+\mathbf{e}_y,\uparrow}^\dag
      -
      c_{\mathbf{j},\uparrow}^\dag
      c_{\mathbf{j}+\mathbf{e}_x-\mathbf{e}_y,\downarrow}^\dag
      +
      c_{\mathbf{j},\downarrow}^\dag
      c_{\mathbf{j}+\mathbf{e}_x-\mathbf{e}_y,\uparrow}^\dag
   \right),
   \\
   \tilde{\Delta}(j_x)
   =&
   \begin{cases}
      \Delta_0,
      \\
      \Delta_0
      \tanh[(j_x-1)/100],
   \end{cases}
   \label{eq:spatila_dep_gap}
\end{align}
In Eq.~(\ref{eq:spatila_dep_gap}), we take into account the fact that 
the coherence length is about $100$ site ($\Delta_0/t=0.01$).
Here we impose impurity potential $V_\mathbf{j}$ for $1\leq j_x\leq N_\mathrm{d}=20$.
\begin{figure*}[htbp]
   \centering
   \includegraphics[width = 17cm]{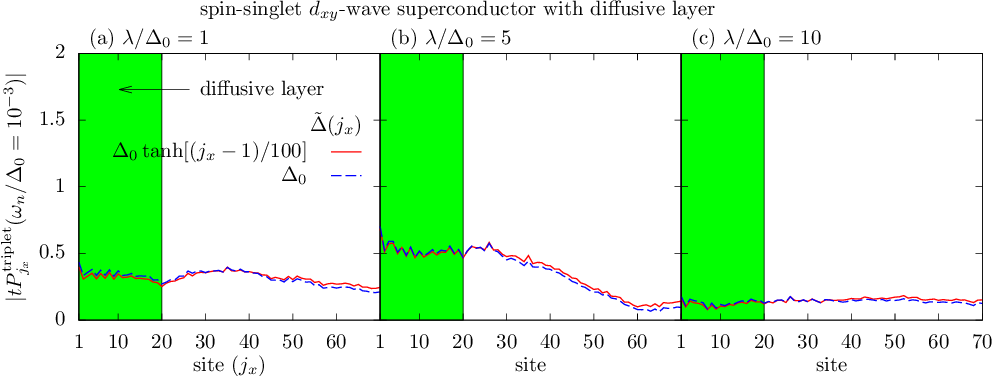}
   \caption{%
      The absolute value of the odd-frequency spin-triplet $s$-wave pair amplitude
      is plotted as a function of site ($j_x$) for 
      the spin-singlet $d_{xy}$-wave superconductor with diffusive layer (green colored region).
      $P^\mathrm{triplet}(\omega_n/\Delta_0=10^{-3})$ for 
      $\tilde{\Delta}(j_x)=\Delta_0$ and $\Delta_0\tanh[(j_x-1)/100]$
      are shown for 
      (a) $\lambda/\Delta_0=1$, 
      (b) $\lambda/\Delta_0=5$ and
      (c) $\lambda/\Delta_0=10$.
   }%
   \label{fig:tanh20}
\end{figure*}
In Fig.~\ref{fig:tanh20}, absolute value of the odd-frequency spin-triplet 
$s$-wave pair amplitude given by
\begin{align}
   |P^\mathrm{triplet}_{j_x}(\omega_n)|
   =&
   \sum_{S_z=1,0,-1}
   \left|
   P^\mathrm{triplet}_{j_x,S_z}(\omega_n)
   \right|,
\end{align}
is shown for $\lambda/\Delta_0=1$, $5$ and $10$. 
Here $P^\mathrm{triplet}_{j_x,S_z}(\omega_n)$ is given by Eqs.~(\ref{eq:def_p_s1}) to~(\ref{eq:def_p_s_1}).
From Fig.~\ref{fig:tanh20}, we can conclude that the spatial dependence of the 
pair potential near the surface almost does not
affect the 
magnitude of the odd-frequency spin-triplet $s$-wave pair amplitude
as compared to the case with constant pair potential.

\end{widetext}
\end{document}